\documentclass[aps,prd,twocolumn,nofootinbib,showpacs,superscriptaddress]{revtex4-1}

\usepackage{amsmath}
\usepackage{amssymb}
\usepackage{slashed}
\usepackage{verbatim} %comment
\usepackage{graphicx}
\usepackage{color}
\usepackage{natbib}

\usepackage{hyperref}

\newcommand{\eq}[1]{eq.~(\ref{#1})}
\newcommand{\Eq}[1]{Eq.~(\ref{#1})}
\newcommand{\tab}[1]{Table~\ref{#1}}
\newcommand{\app}[1]{Appendix~\ref{#1}}
\newcommand{\Ref}[1]{Ref.~\cite{#1}}
\newcommand{\Refs}[1]{Refs.~\cite{#1}}
\newcommand{\fig}[1]{Figure~\ref{#1}}

\newcommand{\ckm}{ {\mathrm{CKM}} }
\newcommand{\qcd}{ {\mathrm{QCD}} }
\newcommand{\Heff}{\mathcal H_\mathrm{eff}}
\newcommand{\mw}{m_\mathrm W}
\newcommand{\VmA}{{\mathrm V - \mathrm A}}
\newcommand{\GF}{G_\mathrm F}

\newcommand{\Sp}[1]{S(#1)}
\newcommand{\Tr}{ \mathrm{Tr}~ }
\newcommand{\qsl}{\slashed{q}}

\newcommand{\sm}{ {\mathrm{SM}} }
\newcommand{\GeV}{\mathrm{GeV}}
\newcommand{\MeV}{\mathrm{MeV}}

\newcommand{\ri}{{ \mathrm{RI} }}
\newcommand{\tree}{{ \mathrm{tree} }}
\newcommand{\lat}{ {\mathrm{lat}} }
\newcommand{\ms}{{\overline{\mathrm{MS}}}}

\newcommand{\pcut}{p_\mathrm{cut}}

\begin{document}

\title{Towards a non-perturbative calculation of Weak Hamiltonian Wilson coefficients}

\author{Mattia~Bruno}
\affiliation{Physics Department, Brookhaven National Laboratory, Upton, NY 11973, USA}

\author{Christoph~Lehner}
\affiliation{Physics Department, Brookhaven National Laboratory, Upton, NY 11973, USA}

\author{Amarjit~Soni}
\affiliation{Physics Department, Brookhaven National Laboratory, Upton, NY 11973, USA}

\collaboration{RBC and UKQCD Collaborations}

\begin{abstract}
We propose a method to compute the Wilson coefficients 
of the weak effective Hamiltonian 
to all orders in the strong coupling constant using 
Lattice QCD simulations. 
We perform our calculations adopting
an unphysically light weak boson mass of around $2~\GeV$.
We demonstrate that systematic errors for
the Wilson coefficients
$C_1$ and $C_2$, related to the current-current four-quark operators,
can be controlled and present a path towards
precise determinations in subsequent works.
\end{abstract}

\maketitle

\section{Introduction}

Weak decays of hadrons, and in particular of mesons, 
play an important role in our understanding of the fundamental forces 
and having precise theoretical predictions to compare against the experimental results
can either strenghten the solidity of the Standard Model or lead to discoveries of new 
physics~\cite{Buras:2016hkx}.

The large scale separation between the mesons, strongly bounded particles with 
masses of order $\Lambda_\qcd$, and the weak mediators with masses around 100 GeV, 
is used to simplify theoretical predictions 
of these processes in the framework
of effective field theories.
By integrating out the heavier degrees of freedom, specifically the weak bosons and 
heavy quarks, from the Standard Model, it is possible to define a new effective Hamiltonian
with new operators and new coupling constants 
usually called the Wilson coefficients.

The coefficients capture the effect of the weak bosons and heavy quarks 
that are absent from the 
effective field theory (EFT), making them well-suited for a perturbative calculation.
Instead, matrix elements of the operators involving mesonic external states require a 
non-perturbative calculation. In the last decade, thanks to algorithmic and computational
advances, the Lattice QCD community has been able to cover a wide range of processes
involving two mesons (see e.g. \Ref{Aoki:2016frl})
and also to complete the first two-body final state decays of $K \to \pi \pi$ 
\cite{Bai:2015nea, Blum:2011ng, Blum:2012uk, Blum:2015ywa}.

On the other hand perturbative calculations of the Wilson coefficients have been successfully carried 
out up to NLO (for a comprehensive review see \Ref{Buchalla:1995vs}) 
and in some cases up to NNLO~\cite{Gorbahn:2004my}.
In this work we explore the possibility of a non-perturbative 
method to compute the Wilson coefficients to 
address the perturbative uncertainty of the analytic calculations.
The perturbative truncation error 
is traded with the statistical and systematic errors usually present in lattice 
calculations
and the purpose of this paper is to define a methodology to obtain a precise determination of the
Wilson coefficients where all uncertainties have been addressed.

The problem of defining the weak effective Hamiltonian non-perturbatively 
has already had some initial considerations by the authors of \Ref{Dawson:1997ic}:
by separating two hadronic weak currents and studying their dependence as 
a function of this distance it is possible to define 
properly normalized operators, where the effect of the Wilson coefficients has been 
fitted away by using their perturbative expansion. 
Our approach differs from the one considered in \Ref{Dawson:1997ic} as we 
plan to directly determine the Wilson coefficients using gauge fixed 
external quark states, rather than mesons, 
in momentum space and not in coordinate space.\\

The rest of the manuscript is organized as follows: in the next
section we give an overview of the main features of the EFT and we describe
our strategy to measure the Wilson coefficients; in section III 
we show our results and address the various uncertanties of the
calculation; in section IV we report our determination of the Wilson coefficients
in the $\ms$ scheme and discuss the comparison against the known perturbative results, 
and finally we conclude and present further directions for this project.

\section{Computational Method}

Before introducing the lattice observables and our main strategy, we 
review the most important features of the EFT.

For concreteness let us restrict to transitions among 
hadrons. When the weak bosons and the heavy quarks are integrated out, 
the leading effective field theory
that arises is based on operators of dimension 6 with four-quark vertices. 
This can be easily seen
by considering the first term in the expasion of the weak propagator
in the limit $\mw \to \infty$.
The full expansion however contains other terms that can be related to 
operators with higher dimensions,
in fact the most general form of the effective Hamiltonian is
\begin{equation}
\Heff = \sum_i V^\ckm_i \frac{\GF}{\sqrt{2}} C_i Q_i 
+ \sum_{j \,, d_j>6} \frac{ c_j^{(d_j)} }{\mw^{d_j-4}} Q_j^{(d_j)} \,.
\label{eq:heff}
\end{equation}
In \eq{eq:heff} $\GF$ represents the dimensionful Fermi constant 
which is related to the $SU(2)_L$ coupling of the Standard Model $g_2$, 
according to $\GF/\sqrt{2} = g_2^2 /(8 \mw^2)$.
$V^\ckm_i$ denotes a generic product of the usual CKM matrix elements that depends 
on the flavor structure of the process and consequently 
of the operators.
Note that among the heavy particles that are removed from the theory 
there is also the top quark, whose presence affects the
Wilson coefficients~\cite{Buchalla:1995vs,Gorbahn:2004my}.

A reliable computation of the Wilson coefficients $C_i$ 
requires a good suppression of the higher dimensional terms, 
which is obtained by performing the calculation
at energy scales sufficiently small compared to $\mw$.
Once the operators in the second sum of \eq{eq:heff} can be neglected, the Wilson coefficients 
are obtained by equating the amplitudes computed in the EFT with the
the ones computed in the full theory, which in our case
differs from the Standard Model as explained below. 
Previous perturbative calculations~\cite{Buchalla:1995vs}
were based on amplitudes computed with off-shell 
external massless quarks and in our calculation
we proceed along these lines.
As ultraviolet quantities, the Wilson coefficients
are expected to be independent from the infrared regulators
of the theory and 
the external states used in the amplitudes;
checking this explicitly will be an important task of our work.
Similarly we will also test the gauge invariance of our results
in the limit where the amplitudes go on-shell,
where any depedence on the weak gauge fixing parameter disappears.\\

The amplitudes of the EFT require additional renormalization conditions, 
due to the new divergences that appear,
leading to the mixing of the renormalized operators among themselves. Hence, it is important
to consider a complete basis of operators that closes under renormalization, such that
\begin{equation}
Q_i^\mathrm R(\mu) = [ Z(\mu) ]_{ij} \ Q_j^\mathrm{bare} \,. 
\end{equation}
The basis of operators depends on the details of the process considered. 
For instance, a $2 \to 2$ transition 
where the four external quarks have different flavors (e.g. $c \to s u \bar d$)
requires only two current-current operators. 
Instead, e.g., a basis of 7 independent operators is needed 
in the 3-flavor EFT (i.e. $\Heff^{\Delta S=1}$) to describe 
the $K \to \pi \pi$ process 
in the zero isospin channel,
involving also disconnected contributions that are 
typically more difficult to compute on the lattice.
Therefore, in this exploratory study we will focus only 
on the simpler current-current operators 
and consequently only on the Wilson coefficients $C_1$ and $C_2$.\\

A third remark concerns the running of the Wilson coefficients, 
as some care is required in computing them at low scales.
In fact, if we naively match the two theories
at scales $\mu \ll \mw$ we may encounter large
logarithms in the form of $\log (\mw^2 / \mu^2)$. 
Therefore, in order to cancel these terms, 
the matching is performed exactly at $\mu=\mw$, thus defining the 
so-called initial conditions of the Wilson coefficients, 
and their values at scales lower than $\mw$ 
are obtained by solving their corresponding renormalization group
equations. 
This leads to a resummation of all logarithms of the form 
$\alpha^n \big( \alpha \log \frac{\mw^2}{\mu^2} \big)^k$ for any power $k$
at fixed loop order $n$.
The step-scaling matrix involved in the running of the Wilson coefficients
is given by the ratio of $Z$ at two different scales and is 
a well known and studied problem on the lattice 
(see e.g. \Refs{Boyle:2017skn,Papinutto:2016xpq}).
Instead, in our work we focus on the 
initial conditions of the Wilson coefficients.
More precisely, we compute the matching between a theory with
3 light dynamical quarks in the sea, playing the role of our 3-flavor EFT, 
with the full theory where we also include the $W$ boson exchange.
In this study we do not consider the problem of removing 
a heavy quark from the theory, which is relevant 
to match the Standard Model, with a top quark, 
onto an EFT with 5 dynamical quarks.\\

Finally, our last remark concerns the feasibility of this 
study. When we introduce the lattice spacing $a$ as a regulator for our theory
we are explicitly introducing an ultraviolet cutoff of order $a^{-1}$.
The Wilson coefficients, in essence, encode the information of momenta
around and above $\mw$, making them potentially very sensitive to discretization
effects. We address this question
by varying both $\mw$ and the lattice spacing.
However, given the current limitations on the availability of fine lattice spacings, we 
perform our calculations in an unphysical scenario, where we take $\mw \approx 2~ \GeV$, 
but where we can control the other systematic uncertainties.
Nevertheless we discuss, before concluding, 
how our results may have an impact 
on the determination of the Wilson coefficients 
with physical values of the weak boson mass.

Once the theory is discretized the path integral can be solved using numerical
simulations, limited by the present computational technologies:
this translates into being able to simulate only finite quark masses and
finite lattice boxes, which constitute the infrared regulators of the theory.
Therefore in order to compute the Wilson coefficients the following limits 
need to be fulfilled
\begin{equation}
m \,, L^{-1} \ll p \ll \mw \ll a^{-1} \,, 
\label{eq:window}
\end{equation}
where $p$ represents the typical momentum of the external states used in the
evaluation of the amplitudes.
If we now consider the limit where $p$ goes to zero, in the infinite-volume 
theory with mass-less quarks, contributions from higher dimensional operators 
should vanish. However, due to dimensional transmutation,
the strong interactions possess a low intrisic scale, $\Lambda_\qcd$,
responsible for the creation of condensates that could
still contribute to \eq{eq:heff} through some operators $Q_j^{(d_j)}$.
Nevertheless in Nature, where $\Lambda_\qcd \ll \mw$, 
these contributions should be suppressed and eventually one may neglect them.
In our exploratory study we achived only $\Lambda_\qcd/\mw \approx 0.2$ 
and in the next sections we describe a strategy to quantify
remaining non-perturbative contaminations and to take them into account in the
systematic uncertanties.\\

In this work we have adopted a momentum subtraction scheme as a prescription
to subtract the divergences of the theory. Alternatively, 
position-space techniques~\cite{Martinelli:1997zc,Gimenez:2004me} could be used in a similar way to directly 
estimate the Wilson coefficients.
In principle the window problem 
sketched in \eq{eq:window} can be partly 
circumvented with finite-size techniques~\cite{Luscher:1991wu}:
in a finite and small box mass-less quarks can be simulated and
the renormalization scale, in our case $p$, can be identified with the box size
itself, thus removing the left-most inequality of \eq{eq:window}. 
In addition to that, very fine lattice
spacings are accessible with present resources 
if the physical volume is small, thus imposing a large
hierarchy between $\mw$ and $\Lambda_\qcd$.
Furthermore, imposing boundary conditions 
on the gauge and quark fields in the temporal
direction, such as Schr\"odinger Functional or Twisted BC~\cite{Luscher:1993gh,deDivitiis:1994yz,Jansen:1995ck,tHooft:1979rtg,tHooft:1981nnx},  
has the additional advantage of producing a well-behaved perturbative expansion.
Calculations with this formalism (see e.g. Refs.~\cite{Bruno:2017gxd,Brida:2016flw}) 
have been carried out successfully for 
different quantities where all systematic uncertainties have been taken into account.
These approaches constitute a valid alternative to further advance 
this project towards a physical value of the weak boson mass.

\subsection{The observables}

As mentioned earlier we restrict ourselves to the simpler current-current operators
$Q_1$ and $Q_2$
\begin{equation}
\begin{split}
Q_1 = & (\bar s_i c_j)_{\VmA} (\bar u_j d_i)_{\VmA} \,, \\
Q_2 = & (\bar s_i c_i)_{\VmA} (\bar u_j d_j)_{\VmA} \,,
\end{split}
\label{eq:Q1_Q2}
\end{equation}
differing only in the color index routing ($i,j$) 
between the two weak currents,
with $ (\bar u_j d_j)_\VmA \equiv \bar u_j \gamma_\mu (1-\gamma_5) d_j $. 
The fields in \eq{eq:Q1_Q2} are understood to reside all at the same space-time point $x$.

\begin{figure}[ht]
\centering
\includegraphics[width=.25\textwidth]{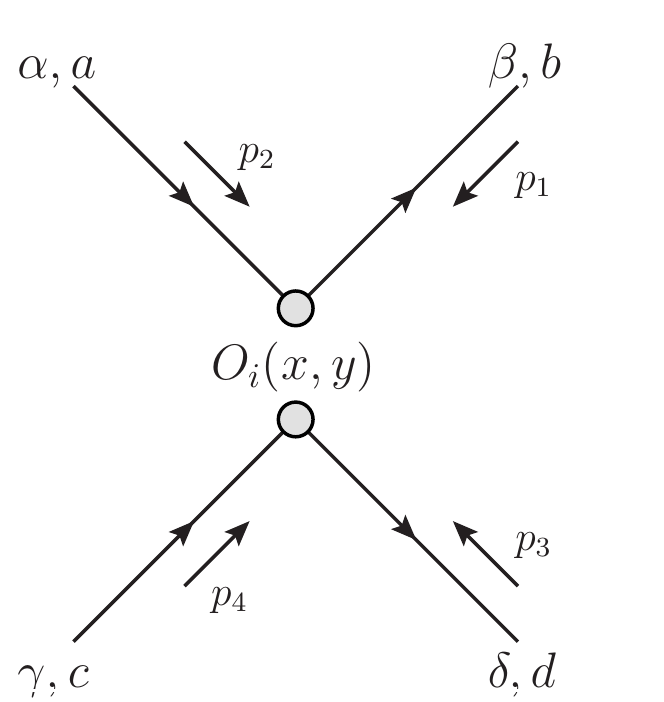}
\caption{Connected diagram of the generic four-point function described in 
\eq{eq:greenfunc} and fourier transformed with phases $e^{ip_k y_k}$. 
Disconnected contributions are forbidden by the flavor structure
of the operators $O_i$. In our notation the four external quarks have all in-coming momenta.}
\label{fig:diagr}
\end{figure}

In the following we introduce a compact notation 
which slightly differs from the one
present in the literature, e.g. \Ref{Buchalla:1995vs}.
Let us define the generic operators $O_i(x,y)$ as
\begin{equation}
\begin{split}
O_1(x,y) = \, & \bar s_i(x) \gamma_\mu^L c_j(x) \ f_{\mu \nu} (x,y) \ \bar u_j(y) \gamma_\nu^L d_i(y) \,, \\
O_2(x,y) = \, & \bar s_i(x) \gamma_\mu^L c_i(x) \ f_{\mu \nu} (x,y) \ \bar u_j(y) \gamma_\nu^L d_j(y)
\end{split}
\label{eq:generic_op}
\end{equation}
with $f$ a generic real-valued function,
$\gamma_\mu^L \equiv \gamma_\mu (1-\gamma_5)$ and
Einstein summation rule on repeated indices.
In this language the choice $f_{\mu \nu} (x,y) = \delta_{\mu \nu} \delta(x-y)$ reproduces
exactly the operators $Q_1$ and $Q_2$ in \eq{eq:Q1_Q2}.
Then let us introduce the following four-point function 
with a single insertion of these operators
\begin{equation}
\begin{split}
[\Gamma(O_i)]^{\alpha \beta \gamma \delta}_{a b c d} & (y_1,y_2,y_3,y_4) = \\
& \langle \, s^\alpha_a(y_2) u^\gamma_c(y_4) \, O_i(x,z) \, \bar c^\beta_b(y_1) \bar d^\delta_d(y_3) \, \rangle 
\end{split}
\label{eq:greenfunc}
\end{equation}
with greek and roman symbols denoting spin and color indices respectively.
By inserting the operators in \eq{eq:generic_op} and computing
the Wick contractions we end up with a Green's function that we later transform
to momentum space, thus obtaining the diagram reported in \fig{fig:diagr}.
In our notation all the four momenta are in-coming.
To simplify the notation we will omit the flavor indices in the rest of the paper 
since we will use degenerate quarks; following the 
subscript $1,2,3,4$ of the $y$ coordinate or of the momenta 
will allow the reader to trace the flavors back.

To compute the quark propagators we invert the Dirac operator $D(x,y)$ on plane 
waves with momentum $p$ (we later refer to them as momentum sources)
\begin{equation}
G(x,p) = \sum_y D^{-1}(x,y) e^{ipy} \,.
\label{eq:Gxp}
\end{equation}
By saturating the $x$-dependence with the appropriate 
phase factor, we obtain the propagator in momentum space and
its expectation value
\begin{align}
\tilde G(x,p) = e^{-ipx} G(x,p) \,, \\
\Sp{p} = \frac{1}{V} \sum_x \langle e^{-ipx} G(x,p) \rangle \,.
\label{eq:props}
\end{align}
The periodicity of the lattice in the temporal and spatial directions
imposes a constraint on the accessible momenta $p_\mu = 2 \pi n_\mu / L_\mu$,
with $n_\mu$ a positive integer. Moreover the breaking of the group
of continuous rotations to the hypercubic ones leads to additional discretization
effects that spoil the smoothness of the propagators as a function of $p$.
To overcome both issues we employ twisted boundary conditions (BC) 
in the valence sector~\cite{deDivitiis:2004kq,Boyle:2003ui,Arthur:2010ht}, 
that allow to access a dense set of momenta\footnote{
With twisted BC we can restrict our calculation to a single irreducible representation
of the hypercubic group, say $ap=(x,-x,0,0)$ with $x \in \mathbb{R}$,
thus obtaining smooth functions.}
\begin{equation}
p_\mu = \frac{2 \pi n_\mu }{L_\mu} + \frac{\theta_\mu}{L_\mu} \,, \quad \theta_\mu \in [-\pi/2, \pi/2] \,.
\label{eq:twistedBC}
\end{equation}

For simplicity we give below the explicit expression of $\Gamma(O_2)$, where
we omit the index contractions inside the square brackets and where
we use the $\gamma_5$-hermiticity of the Dirac operator
\begin{equation}
\begin{split}
[\Gamma(O_2) & ]^{\alpha \beta \gamma \delta}_{a b c d} (p_1,p_2,p_3,p_4) = \sum_{\mu \nu} \sum_{x,y} \\
& \langle \big[ \gamma_5 \tilde G(x,-p_2)^\dagger \gamma_5 \gamma_\mu^L \tilde G(x,p_1) \big]^{\alpha \beta}_{a b} f_{\mu \nu}(x,y)  \\
& \times \big[ \gamma_5 \tilde G(y,-p_4)^\dagger \gamma_5 \gamma_\nu^L \tilde G(y,p_3) \big]^{\gamma \delta}_{c d} \rangle \,.
\end{split}
\label{eq:gamma_O2}
\end{equation}
The amputated amplitudes $\Lambda$ are easily obtained by inverting
the expectation value of quark propagators $S(p_i)$ in \eq{eq:props}
\begin{equation}
\begin{split}
[\Lambda(\Gamma) ]^{\alpha \beta \gamma \delta}_{a b c d} = & 
  [\Sp{p_2}^{-1}]^{\alpha \alpha^\prime}_{a a^\prime}
  [\Sp{p_1}^{-1}]^{\beta \beta^\prime}_{b b^\prime}
  [\Sp{p_4}^{-1}]^{\gamma \gamma^\prime}_{c c^\prime} \\ &
  [\Sp{p_3}^{-1}]^{\delta \delta^\prime}_{d d^\prime}
\Gamma^{\alpha^\prime \beta^\prime \gamma^\prime \delta^\prime}_{a^\prime b^\prime c^\prime d^\prime} \,.
\end{split}
\label{eq:Lambda}
\end{equation}

At this point we define the amputated amplitudes $\Lambda_i \equiv \Lambda(\Gamma(O_i))$ 
with $f_{\mu \nu} (x,y)=\delta_{\mu \nu} \delta(x-y)$.
On the full theory side, only a single operator with color diagonal structure 
is needed to describe this process, namely $O_2$ in \eq{eq:generic_op} 
with $f_{\mu \nu}(x,y) = W_{\mu \nu}(x,y)$, the tree-level propagator 
of the weak charged bosons in coordinate
space (Euclidean metric), which we obtain by fourier 
transforming\footnote{
In our calculation we use the lattice momenta $ a \hat p_\mu = 2 \sin(a p_\mu/2)$ in 
\eq{eq:wprop}.
}
\begin{equation}
W_{\mu \nu}(p) = \frac{1}{p^2 + \mw^2} \Big( \delta_{\mu\nu} - \frac{p_\mu p_\nu}{\mw^2} \Big) \,. 
\label{eq:wprop}
\end{equation}
Eqs.~(\ref{eq:greenfunc},\ref{eq:gamma_O2},\ref{eq:Lambda}) hold in this case as well
and we define $\Lambda_\sm$ as the amputated amplitude obtained from $O_2$ with $f$ replaced
by the $W$ boson propagator given above.
The choice of the unitary gauge simplifies the calculation in the full theory to a
single diagram. In the next section we present results also for the 
Feynman gauge ($R_\xi$ gauge with $\xi=1$) where the contribution of the Goldstone
boson needs to be included. \\

The $\delta(x-y)$ function in the insertion of the local operators $Q_i$ 
simplifies the double sum over $x$ and $y$ in \eq{eq:gamma_O2} to a single one. 
Choosing plane waves at the source of the propagators allows us to perform
such a sum over the entire volume, thus sampling the operators $Q_i$ 
on the full lattice and reducing the statistical fluctuations of the final amplitudes.
When $f$ is replaced by the $W$ boson propagator, the double sum needs to be 
performed. Also in this case the usage of momentum sources gives us the freedom 
(at the sink of the propagators) to evaluate both sums over $x$ and $y$ explicitly,
thus significantly reducing the noise of $\Lambda_\sm$.\\

The only drawback of the combined use of momentum sources and twisted BCs is that
a separate inversion is necessary for each momentum configuration that is considered,
up to a total of 4 per configuration if we choose the four external legs with different
momenta. Therefore, in an effort to balance the cost of the inversions against the benefit of the volume
average, we have explored a second strategy to sample the $W$ boson propagator.
From the simple observation that a point-source propagator can be fourier transformed to any
continuous momentum, up to small finite-size errors,
we have studied the possibility to stochastically sample $W_{\mu\nu}(x,y)$, 
with $x$ and $y$ being the sources of the inversions
rather than the sinks as in the case of momentum sources.
The sampling technique is essentially borrowed from the calculation of the light-by-light contribution to the anomalous
magnetic moment of the muon by the RBC/UKQCD collaboration~\cite{Blum:2015gfa} 
and it proceeds in two steps: 

\begin{itemize}
\item for a fixed $x$, the sum over $y$ in \eq{eq:gamma_O2} is achieved by randomly sampling $W_{\mu \nu}$ 
with a probability distribution falling off rapidly for large separations $|x-y|$; to recover
the flat sum over $y$ the appropriate reweighting factor is applied and to further decrease the cost, 
the hypercubic symmetries are taken into account by randomly sampling only one element per equivalence
class, defined by all the points $y$ with the same distance from $x$, $|x-y|$; this procedure 
defines a cloud of points stochastically sampled around the center $x$;

\item to reduce the noise of the observables a second sum 
over the center of the cloud, $x$, is performed, 
which is again stochastic and with a flat distribution.

\end{itemize}

Even though a continuous set of momenta is now accessible through the usage of the point sources 
(exceptional and non-exceptional kinematics can also be explored simultaneously), 
the statistical noise grows quickly: in our test we have used 40 points inside the cloud and this
led to a controlled approximation of the sum over $y$ (for the different values of the 
input $W$ mass considered in this work); however the second sum over $x$, for which we used
16 different clouds, turns out to be the crucial one in further reducing the noise. 
Although the presence of a finite correlation length in our system
decreases the number of useful points to reduce the noise,
we have verified that the stochastic sampling leads to results that are
at least 10 times noisier compared to the momentum source method, for approximately the same
cost. For this reason we leave all the details of this secondary approach to the \app{app:point-source}
and we concentrate in the rest of the manuscript on the analysis of the results from momentum sources.
Nevertheless this technique has been very useful in an early stage of the work when we explored a
vast range of momenta and kinematic configurations, on which we based our decision for the final strategy.

\subsection{The Wilson coefficients}

The appearance of new divergences in the EFT requires the introduction
of additional renormalization conditions. In our study we adopt a variety
of regularization independent schemes, called RI/MOM and RI/SMOM, 
which were introduced in \Refs{Martinelli:1994ty,Aoki:2007xm,Sturm:2009kb},
that are entirely defined by the choice of the external states and
projectors: 
this translates into the momentum combination used in the calculation of the 
propagators and the projectors that we apply on the amputated Green's
functions to obtain definite spin-color states.

In this paper we have explored two combinations of the 
four momenta on the external legs: the non-exceptional case
called RI/SMOM where $(p_i+p_j)\neq0$ for all pairs $i\neq j$,
and the exceptional one (RI/MOM)
where at least one linear combination of the external momenta vanishes.
If the amplitude under study possesses the same symmetries of the 
light mesons, at momenta comparable with their Compton wave length,
the smooth perturbative behavior of the amplitude may be spoiled by 
non-perturbative contaminations, referred to as Goldstone-pole contaminations~\cite{Martinelli:1994ty,Aoki:2007xm}.
Using non-exceptional kinematics (RI/SMOM) suppresses these unwanted effects~\cite{Aoki:2007xm,Sturm:2009kb}
and we extensively discuss them in the next section.\\

To define specific spin-color states we use two projectors $P_i$ 
($i=1,2$) that we apply to the amputated amplitudes to define the matrix 
\begin{equation}
M_{ij} = \Tr \big( P_j \Lambda_i \big) \,,
\label{eq:M}
\end{equation}
such that $M$ is invertible.
To achieve this we fix the color structure of the projectors to one of the operators
\begin{equation}
P_1 = \delta_{bc} \delta_{da} \, \Gamma_1 \otimes \Gamma_2 \,, \quad
P_2 = \delta_{ba} \delta_{dc} \, \Gamma_1 \otimes \Gamma_2 \,,
\end{equation}
and we explore two options for the Dirac part, 
with different parities (even and odd)
\begin{equation}
\Gamma_1 \otimes \Gamma_2 = \bigg\lbrace
\begin{array}{lcl}
\gamma_\mu \otimes \gamma_\mu & + & \gamma_\mu \gamma_5 \otimes \gamma_\mu \gamma_5 \\
\gamma_\mu \gamma_5 \otimes \gamma_\mu & + & \gamma_\mu \otimes \gamma_\mu \gamma_5 
\end{array}\,.
\label{eq:gammaproj}
\end{equation}
\Eq{eq:gammaproj} defines the so-called $\gamma$ (or $\gamma_\mu$) projectors. 
Alternatively, 
the replacement $\gamma_\mu \to \qsl /|q|$ and $\gamma_\mu \gamma_5 \to \qsl \gamma_5/|q|$
defines the $\qsl$ projectors.
Computing the Wilson coefficients using both 
parities and $\gamma$ or $\qsl$ Dirac structures 
turns out to be a
crucial test of the calculation, as explained in the next section.
In the rest of the paper we will refer to the two parities as
$\mathrm{VV}+\mathrm{AA}$ and $\mathrm{VA}+\mathrm{AV}$, with 
$V$ and $A$ labeling vector and axial Dirac structures.\\

The renormalization conditions that we impose on the four-quark
operators read as follows
\begin{equation}
\lim_{m_q \to 0} \ 
\frac{Z_{ij}^\ri}{(Z_q^\ri)^2} \, 
M_{jk}^\lat|_{\mu^2=p^2} = M_{ik}^\ri |_{\mu^2=p^2} \equiv M_{ik}^\tree
\label{eq:rimom}
\end{equation}
with
\begin{equation}
Z_q^\ri \Sp{p}^\lat = \Sp{p}^\ri \,,
\end{equation}
$^\lat$ indicating bare lattice quantities
and $M_\tree$ defined by replacing the amplitudes in \eq{eq:M} 
with their tree-level counterparts.

In the full theory no additional renormalization conditions are required
for the projected amplitude, which we denote with $W_i = \Tr ( P_i \Lambda_\sm )$, 
beyond the usual wave function renormalization introduced above.
It is important to note that since we are considering the weak theory at tree level, 
$\mw$ does not renormalize, since self-energy diagrams of the
$W$ boson propagator appear at higher orders in
the weak coupling constant, $g_2$.
In the continuum theory, vector and axial 
current conservations (for mass-less quarks) guarantee that
the same is true for $g_2$ as well.
On the lattice however the usage of 
local vector and axial currents 
dictates the presence of the finite
renormalization factors $Z_V$ and $Z_A$, 
thus leading to $g_2^R = Z_V g_2$.
Note that the $V-A$ current can be renormalized with either $Z_A$ or $Z_V$
due to the excellent chiral properties of the Domain Wall formulation
and to the employment of non-exceptional kinematics, as described later.
Eventually we opt for $Z_V$ to avoid additional 
chiral symmetry breaking effects and a larger quark mass
dependence compared to $Z_A$.\\

Now we have all the basic ingredients to impose the matching between the two theories 
\begin{equation}
\frac{\GF}{\sqrt{2}} \ C_i^{\ri}(\mu) M_{ij}^{\ri}(\mu) = W_j^{\ri}(\mu) =
\frac{g_2^2}{8} \frac{Z_V^2}{(Z_q^\ri)^2} W_j^\lat \,.
\label{eq:matching}
\end{equation}
We have already simplified the usual CKM factors
that appear in the same form on both sides of the equation. 
The weak coupling constant $g_2$ 
simplifies as well with the Fermi constant $\GF$, leaving only a factor $\mw^2$ on the right hand side.

By expanding $M^\ri$ and $W^\ri$ with their corresponding bare lattice counterparts we 
obtain the definition of the Wilson coefficients
\begin{equation}
\begin{split}
C_i^\lat \equiv & \ \mw^2 \Big( W_j^{\lat} [M^{\lat}]^{-1}_{ji} \Big) \,, \\
C_i^{\ri}(\mu) = & \ C_j^\lat  \
 \Big( [Z^{\ri}(\mu)]^{-1}_{ji} Z_V^2 \Big) \,.
 \end{split}
\label{eq:wcoeff}
\end{equation}
\Eq{eq:wcoeff} nicely separates two basic ingredients and consequently 
two different problems: 

\begin{itemize}

\item the bare lattice Wilson coefficients $C_i^\lat$
which can be used to inspect the size of the higher dimensional operators and other systematic errors, 
and that we compute at small momenta for a variety of external states;

\item the renormalization factors that we compute 
at high momenta and use to renormalize $C_i^\lat$ and eventually connect to the 
$\ms$ scheme, by means of the one-loop conversion matrix $Z^{\ri \to \ms}$ given in 
the \app{app:rims}.

\end{itemize}

Our analysis proceeds along these two steps: we first examine the dependence of 
the bare Wilson coefficients on the momentum scale, the quark mass and the finite box size.
Then we briefly describe the results for the renormalization matrices 
and discuss discretization errors, 
and finally present the comparison against the known perturbative results
in the $\ms$ scheme.

\section{Results}

In our study we have used 
the Domain Wall formalism, 
which retains excellent chiral properties, even at finite 
lattice spacing~\cite{Blum:1996jf,Blum:1997mz} %with $O(10)$ sites in the fifth dimension~\cite{Blum:1996jf,Blum:1997mz}
(which become exact in the limit of
infinite 5th dimension~\cite{Shamir:1993zy,Furman:1994ky}), 
simplifying the renormalization pattern of the theory and 
suppressing the mixing among operators belonging to different representations of the chiral group~\cite{Blum:2001xb}.

We have measured the amplitudes described in the previous section on three different ensembles, 
reported in \tab{tab:ensembles}, with 2+1 Domain Wall Fermions (DWF) in the sea. 
We have used a unitary setup by promoting the same discretization (Shamir DWF)
also to the valence sector. Ensembles 16I and 24I have been used to study the
dependence of the Wilson coefficients on the volume of the lattice. The additional ensemble 32I 
allows us to take the continuum limit, with two lattice spacings differing approximately by
a factor of 2 in $a^2$. In all our calculations we have fixed the (QCD) gauge to the Landau gauge.\\

\begin{table}
\begin{tabular}{ccccccc}
\toprule
Id & $\beta$ & $a~[\GeV^{-1}]$ & $L/a \times T/a$ & $am_\mathrm{u,d}$ & $am_\mathrm s$\\
%\midrule
\colrule
16I & 2.13 & 1.78 & $16^3 \times 32$ & 0.01  &  0.04 \\
24I & 2.13 & 1.78 & $24^3 \times 64$ & 0.01  &  0.04 \\
32I & 2.25 & 2.38 & $32^3 \times 64$ & 0.008 &  0.03 \\ 
\botrule
\end{tabular}
\caption{List of the ensembles used in this work. In the table we report the most important
physical parameters, the remaining details can be found in \Refs{Aoki:2010dy,Allton:2007hx}. 
The three ensembles have been generated with
the same Domain Wall discretization for the sea quarks, 
the Shamir formulation with  
fifth dimension length $L_s = 16$, 
and Iwasaki gauge action, 
with bare couplings $g_0^2= 6 / \beta$ reported in the second column.
In the third column we quote the lattice spacings measured in \Ref{Blum:2014tka} 
and in the last ones we provide the lattice dimensions and the bare sea quark masses.
}
\label{tab:ensembles}
\end{table}

Our measurements require the calculation of quark propagators for which we have used a 
mass (in lattice units) of 0.04 on the 24I and 16I ensemble, and 0.03 on the 32I.
We utilize 27 independent configurations for both 24I and 32I, 
separated by 100 and 200 Molecular Dynamics Units.
In our analysis we use the jackknife method with bin size of 1,
after checking the stability of the error of the Wilson coefficients
for larger bin sizes.
On our coarser ensembles, the 16I and 24I, we measure our propagators
up to momenta of $O(0.8~\GeV)$, with data points evenly spaced in $p^2$.
On each configuration we perform 4 different inversions at fixed $p^2$ for the
four different legs required in non-exceptional kinematics.
Moreover on the 16I and 24I we also measure the same amplitudes with 
momentum injected explicitly along the time direction to test finite volume 
effects as explained later. On the 24I we repeat those measurements for three
values of the quark mass, with $am=0.02\,, 0.04$ and 0.08.
Finally on the 32I we compute the amplitudes for four different momenta
up to $0.4~\GeV$.
Our calculations of $\Lambda_\sm$ cover a range for $a \mw$ 
that goes from $0.6$ to $1.334$ on the 24I and
$0.6 < a \mw < 1.0$ on the 32I: the masses on the 32I have been tuned to match those
on the 24I according to the ratio of lattice spacings computed in \Ref{Blum:2014tka}.\\

When the momentum of the external quark states becomes comparable to $\mw$, the four-quark EFT
is expected to deviate from the full theory, due to the lack of higher dimensional operators
that eventually become relevant. Therefore for different choices of the external states in our definition
of the amplitudes, we expect to observe different behaviors with respect to the dominant scale $p^2$. 
However, in the limit where $p^2 / \mw^2 \ll 0$ they should all agree and give a consistent
and unique value for the Wilson coefficients, up to $O(\Lambda_\qcd)$ contributions. 
This suggests that we should be able to fit $C_i^\lat$ with a polynomial 
function of $p^2/\mw^2$ and we turn to this now.
In the next sections we address 
the problems related to the usage of small momenta 
such as finite volume and finite quark mass effects
and possible remaining
non-perturbative effects of order $\Lambda_\qcd / \mw$.

\subsection{Fitting strategy}

In order to address the size of higher dimensional operators
we study the momentum dependence of the $C_i^\lat$s for several choices of the
external states: we adopt a non-exceptional momentum configuration given
by
\begin{equation}
\begin{split}
p_1 = (x,-x,0,0) \quad p_2 = (0,0,-x,x) \\
p_3 = (-x,0,x,0) \quad p_4 = (0,x,0,-x)
\end{split}
\label{eq:nonex}
\end{equation}
with $p_i^2 = p^2$, $\forall i$, and transfer momentum $q^2 =2 p^2$.
In \eq{eq:nonex} $x$ denotes a continuous parameter obtained
according to \eq{eq:twistedBC}.
The exceptional momentum configuration is easily obtained 
by re-using the same propagator with one of the four momenta in \eq{eq:nonex}
on all the four legs. The momentum configuration proposed
in \eq{eq:nonex} leads to momentum conservation in the amplitudes and 
to exact equivalence between $\gamma$ and $\qsl$ projectors.\\

\begin{figure}[!htb]
\includegraphics[width=.49\textwidth]{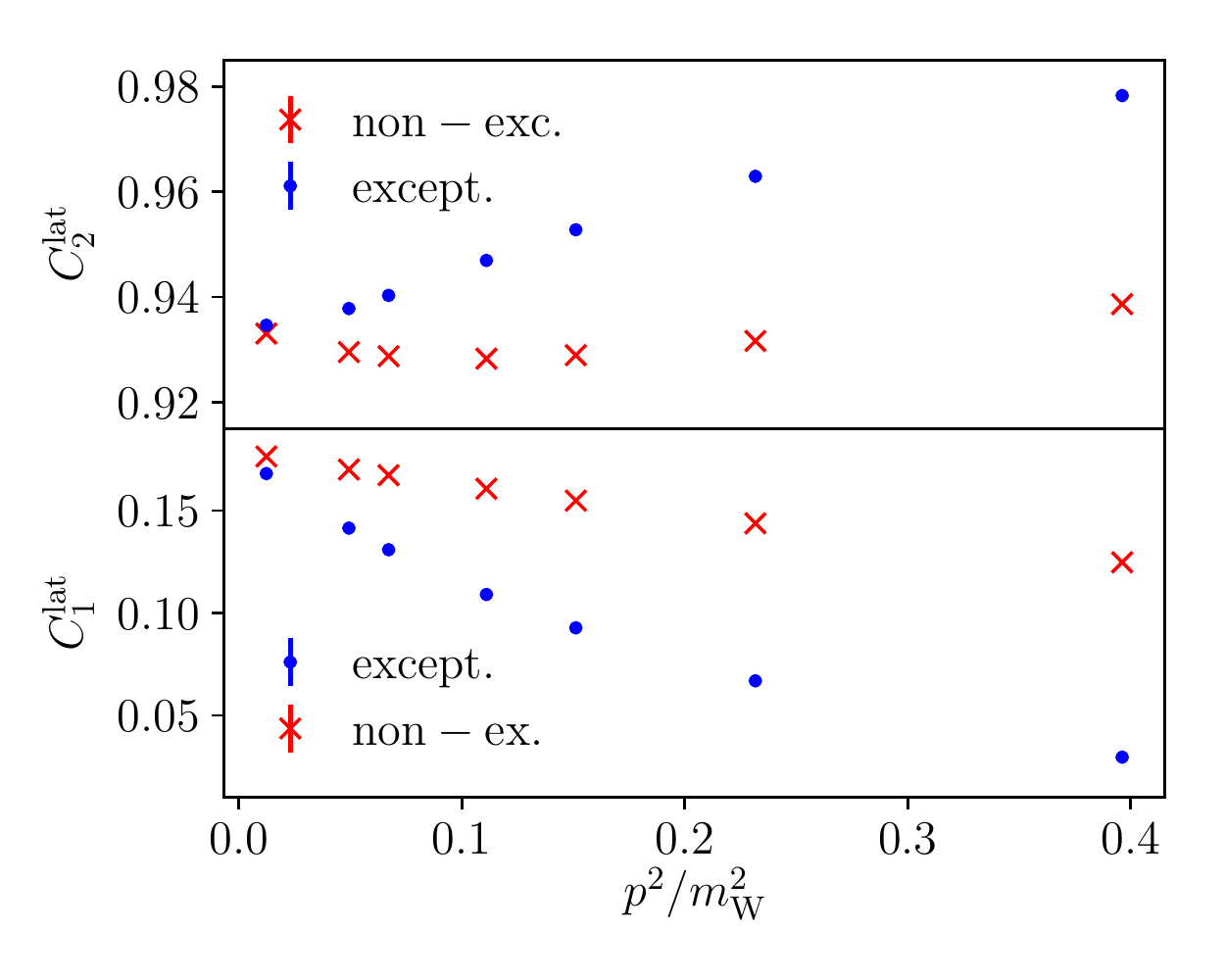}
\caption{Dependence of the bare lattice Wilson coefficients 
on the momentum of the external quark states 
for a fixed value of $\mw \approx 1.8~\GeV$.
The two sets of data points correspond to the exceptional and 
non-exceptional kinematics described in \eq{eq:nonex}
measured on the 24I ensemble with $am=0.04$.
We perform combined polynomial fits where the constant term
is constrained by the two data sets, which have been
obtained from $\qsl$ parity odd projectors.}
\label{fig:c12lat}
\end{figure}

In \fig{fig:c12lat} we show the results of $C_1^\lat$ and 
$C_2^\lat$ as a function of $p^2/\mw^2$ 
from the 24I ensemble. We observe a good convergence of the 
two sets of data points (exceptional and non-exceptional) 
for small values of the expansion variable $p^2/\mw^2$.
Nevertheless higher dimensional operators are quite sizable, 
given the high accuracy that we are able to achieve, 
which forces us to explore a particularly small range of momenta and eventually
consider extrapolations to $p^2/\mw^2 \to 0$.
However, data obtained with non-exceptional kinematics show a milder dependence
on the external momentum $p^2$.

To extract the bare values of the Wilson coefficients showed 
in \fig{fig:c12lat}, we adopt combined polynomial fits 
to the exceptional and non-exceptional points with a common 
constant term
\begin{equation}
\begin{split}
C_i^{\lat,\mathrm{ex}} (x)     & = C_i^\lat + \sum_{k=1}^N  A_{ik}  x^k \,, \quad i=1\,,2 \\ 
C_i^{\lat,\mathrm{nonex}}  (x) & = C_i^\lat + \sum_{k=1}^N  B_{ik}  x^k \,, \quad x=\frac{p^2}{\mw^2} \,. 
\end{split}
\label{eq:fits}
\end{equation}

We base our decision for a combined fit 
on the fact that independent fits to the two data sets
reproduce results for $C_i^\lat$ well 
compatible within 1 standard deviation (on a given fit range).
To estimate the systematic uncertainties associated with these extrapolations
we vary the upper limit of the fit range and the degree of the polynomial ($N=1,2,3$) 
and we consider
the fits with good $\chi^2$ per d.o.f:
we take as our final value the result from the highest polynomial,
whose larger statistical error covers the discrepancy 
among the several 
fits\footnote{
This excludes only linear fits 
with $ap_\mathrm{max} > 0.15$ on the 24I, the ensemble for which we have a wider
number of measurements. In the 32I case all extrapolations are compatible with each other.
}.

With the lattice spacings that we have studied in this work, higher dimensional operators
seems to be sufficiently suppressed only for momenta around and below $\Lambda_\qcd$.
Therefore studying the dependence on the infrared regulators of the
lattice Wilson coefficients is important to control their extraction.
In general we expect them to be small in $C_i^\lat$ due to their 
cancellation between $W$ and $M$, 
especially if the quark momentum is well below $\mw$.
However to what degree this is realized in practical 
non-perturbative simulations 
is not clear a priori and we study that below.

\subsection{Finite Volume Errors}

\begin{figure*}[!htb]
\includegraphics[width=0.48\textwidth]{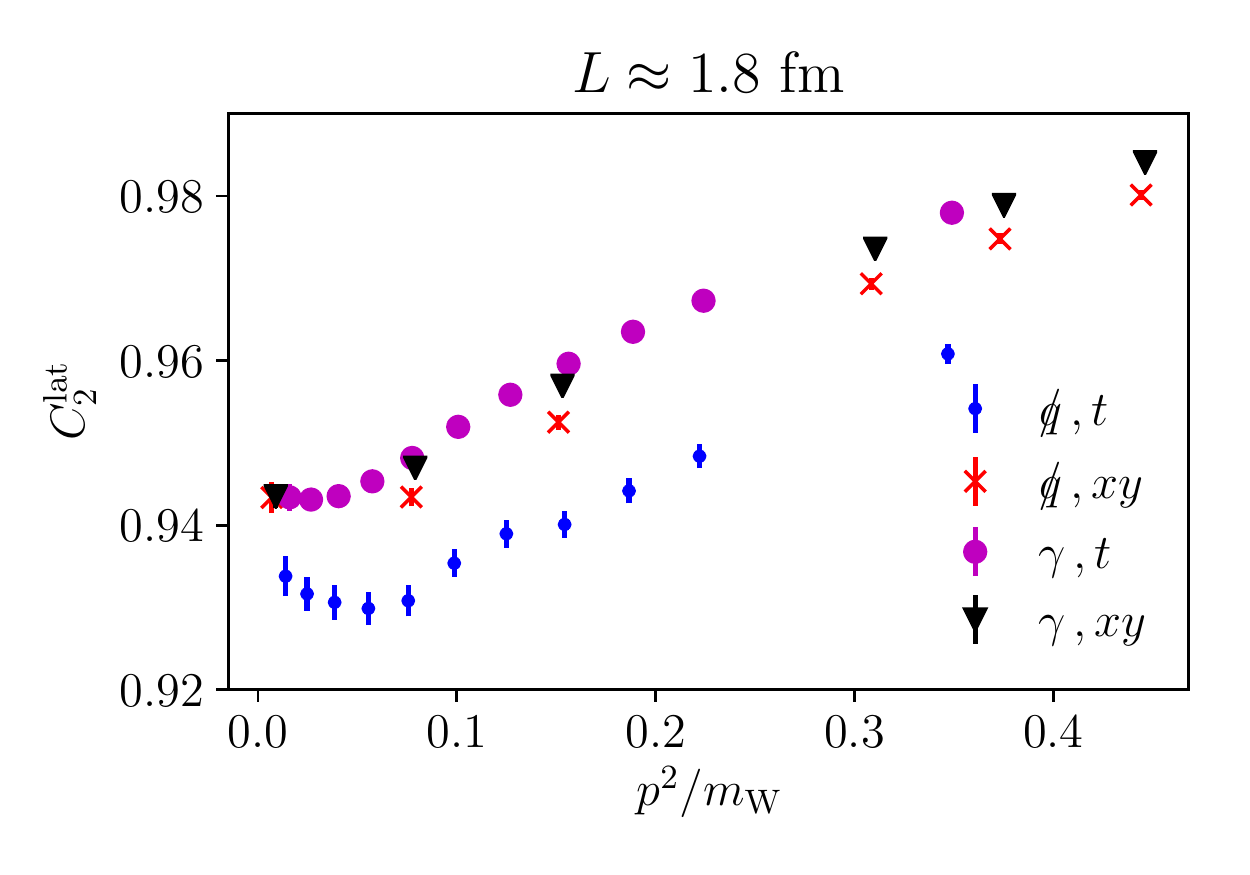}
\includegraphics[width=0.48\textwidth]{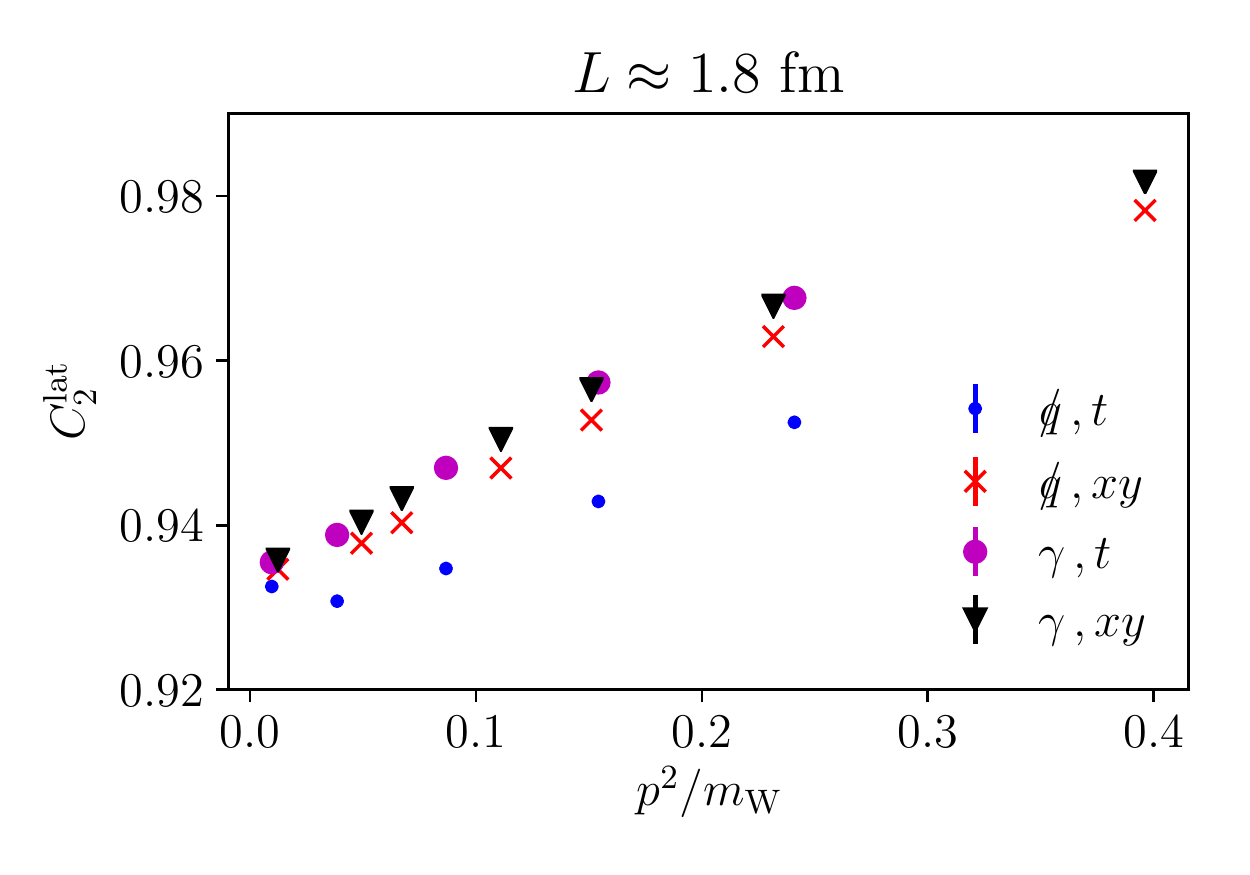}
\caption{Dependence of the lattice Wilson coefficient $C_2^\lat$ on the 
momentum of the external states for fixed $\mw\approx 1.8~\GeV$.
The labels in the legend denote the choice of projectors ($\gamma$ or $\slashed{q}$) 
and momentum (injected along $xy$ spatial directions or time $t$).
Both panels show points obtained from parity odd projectors.
{\it Left}: results from the 16I ensemble; if the momentum is along
the time direction, twice as big as the spatial ones, $C_2^\lat$ from $\slashed{q}$ projectors
coincides with the measurements performed on the larger lattice. {\it Right}: 
results from the 24I ensemble where all combinations of momentum and projectors
converge to the same point in the limit of small momenta.} 
\label{fig:finite_volume}
\end{figure*}

The first infrared regulator that we investigate is the box size.
For this purpose we have measured the lattice Wilson coefficients
on the 16I and 24I ensembles using exceptional kinematics. 
In \fig{fig:finite_volume} we present the results for $C_2^\lat$: 
note that the two plots, corresponding to the two ensembles, share the same
y-axis to facilitate a visual comparison.
Both plots contain four sets of points obtained by combining $\gamma$ and 
$\slashed{q}$ projectors with amplitudes where the momentum is injected 
along $xy$ and time directions. The parity of the projectors corresponds
to $\mathrm{VA}+\mathrm{AV}$.

For the 24I ensemble the various measurements converge to a unique point 
at small momenta, thus remarking the universality of the Wilson coefficients
in that limit. However for the 16I ensemble a different behavior is observed: in this
case only the combination of $\slashed{q}$ projectors with amplitudes with
momentum along the time direction agrees with the correponding measurements in
the larger volume, while the other sets of points converge to a different value.
We interpret such a behavior as a finite volume error, 
that is largely suppressed when the momentum
is injected along the time direction, which is twice as big as the spatial ones.
From our numerical analysis we have noticed that 
the finite volume effect measured in the 16I ensemble decreases 
with $\mw^{-1}$. In agreement with other observations presented later,
 this may be a consequence of a dimension-7 operator involving 
a non-perturbative condensate, whose strong dependence on the box volume
eventually generates the spread that we observed.

A similar behavior is observed also in the other Wilson coefficient, $C_1^\lat$, 
where the measurements on the 16I do not agree at small momenta, in constrast with 
the 24I ensemble.

\subsection{Non-perturbative effects}

In our study we have not been able
to achieve a large separation between the strong and weak scales, 
since $\Lambda_\qcd / \mw \approx 0.2$. 
This means that our calculation, based on RI/MOM techniques,
might potentially suffer from non-perturbative contaminations
often referred to as Goldstone-pole contaminations~\cite{Martinelli:1994ty,Aoki:2007xm}:
in general varying the quark mass in the measurements of the amplitudes
is useful to address this issue and has been previously used 
to non-perturbatively subtract them away~\cite{Becirevic:2004ny}.
Such contaminations are present mostly 
in observables that share the same quantum numbers of the lighter mesons,
such as the axial or pseudo-scalar bilinear operators (see for example \Ref{Giusti:2000jr}).
If we extend the discussion to fermionic formulations 
that do not preserve chiral symmetry, such as Wilson fermions, 
also the mixing with wrong chiralities is allowed and at small
momenta can lead to pole behavior as well. For our study we have checked
how well chiral symmetry is realized by repeating some measurements of the 
Wilson coefficients with a larger separation of the Domain Walls along 
the fifth dimension: we have obtained an excellent agreement for all combinations
of projectors between results obtained with the Shamir formulation with $L_s=32$
and $L_s=16$ (the latter being the same one used for the sea quarks).
Hence, we expect any mixing with wrong chiralities to be predominantly 
of infrared origin, due to the spontaneous breaking of chiral symmetry 
via quark condensates, and to vanish at high momenta.
Based on $\mathcal{CPS}$ symmetry arguments~\cite{Bernard:1985wf,Bernard:1987pr,Donini:1999sf}, 
we also expect the 
parity odd projectors to show less contaminations compared 
to the parity even case.

\begin{figure}[htb]
\includegraphics[width=.49\textwidth]{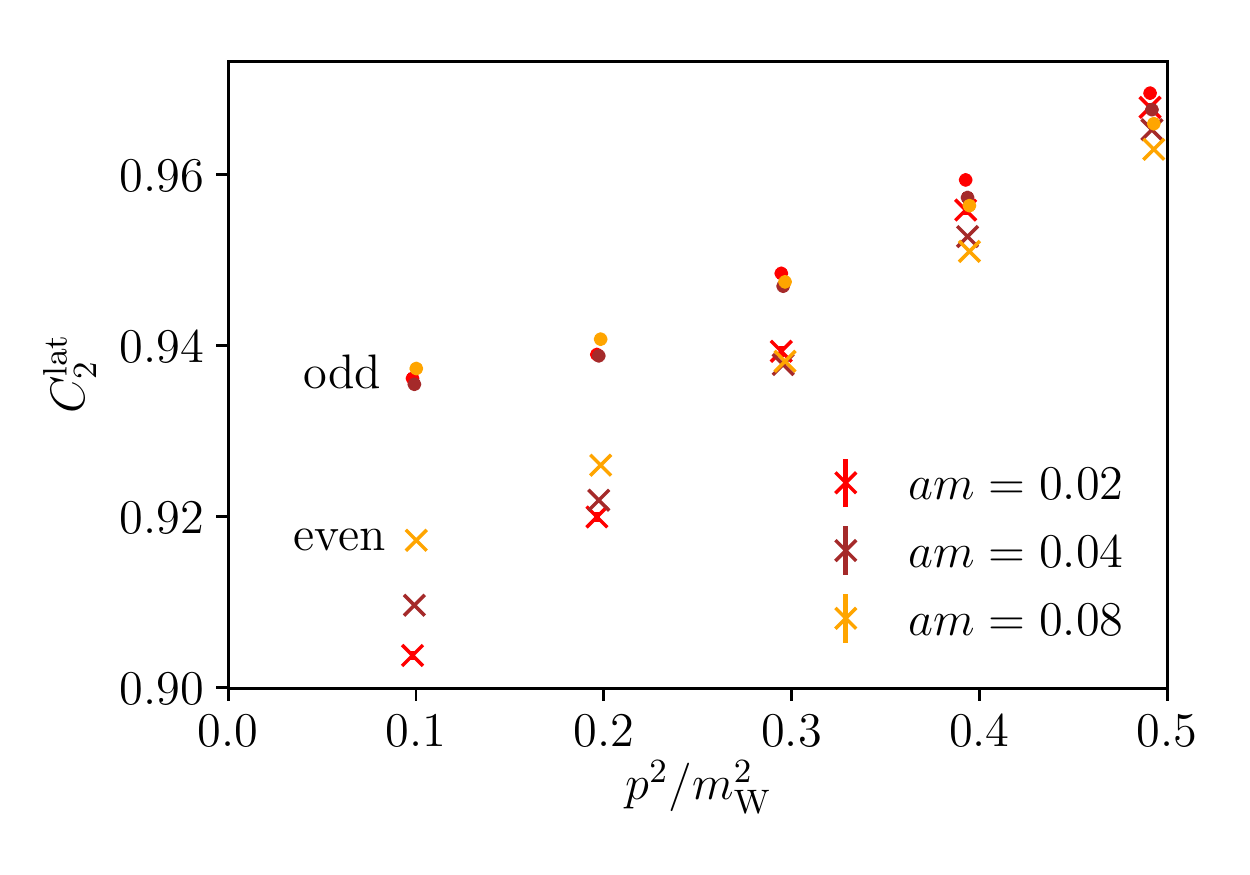}
\caption{Quark mass dependence of $C_2^\lat$ for fixed input $W$ mass at $1.8~\GeV$ on the 24I 
ensemble. Crosses refer to parity even projectors 
and dots represent the opposite parity, all in the $\gamma$ scheme. 
No difference is observed if $\gamma$ projectors are replaced with the corresponding $\qsl$ ones.
The $\mathrm{VV+AA}$ projectors produce results very sensitive to the quark mass, 
contaminated by infrared effects proportional to $1/p^2$. 
A small quark mass dependence is observed
in $C_1^\lat$ for parity odd projectors as well.}
\label{fig:quarkmass}
\end{figure}

To quantify this problem we start from 
the difference between the Wilson coefficients computed with parity even and 
parity odd projectors, defined in \eq{eq:gammaproj}, for several values of the valence 
quark mass.
In \fig{fig:quarkmass} we show the results for the lattice Wilson coefficient $C_2^\lat$ 
from the exceptional momentum configuration, where we vary the quark mass and
the parity of the ($\gamma$) projectors used to compute $M^\lat_{ij}$ and $W_i^\lat$. 
For parity odd projectors we observe an excellent agreement among the 
different quark masses for all points, down to the smallest momentum.
On the other hand parity even projectors lead to a strong dependence of $C_2^\lat$
on the quark mass,
well compatible with the expectation of a 
Goldstone-pole contamination: in fact, by increasing the quark mass 
from 0.02 to 0.08 the data is driven towards the points obtained 
from parity odd projectors, due to the suppression
of non-perturbative effects which we expect to be 
proportional to $(p^2 + m^2)^{-1}$, with $m$ the mass of
the light state coupling to the amplitude.
Changing from $\gamma$ to $\qsl$ projectors does not affect the general trend
showed in \fig{fig:quarkmass}, as well as turning to the non-exceptional kinematics
in \eq{eq:nonex}. 
For $C_1^\lat$ we draw similar conclusions on the behavior of the parity even results, 
with the only exception that a quark mass depedence is visible 
also for parity odd projectors: given the 
different nature and size of this Wilson coefficient it turns out to be large, 
on the 10\% level, but only 0.015 in absolute units (with $C_1^\lat \approx 0.15$). 
We account for this effect
in our systematic error.\\

To better understand the origin of the discrepancy 
between the two parities,
we have projected our amputated Green's functions 
on the so-called wrong chiralities.
For odd parity no significant mixing has been found.
Instead we have measured strong
contributions at small momenta from the projectors
with  form $1 \otimes 1$, $\gamma_5 \otimes \gamma_5$ and
$\sigma_{\mu \nu} \otimes \sigma_{\mu \nu}$, which
quickly vanish above $1~\GeV$.
This provides another indication on the pollution of infrared
effects in the parity even sector, 
as expected from $\mathcal{CPS}$ symmetry.
In \app{app:subproj} we describe an alternative approach
consisting of changing the definition of the projectors
to suppress the overlap with the unwanted chiralities.\\

Checking the gauge independence of our calculation is a second
approach that we exploit to quantify 
non-perturbative contributions of $O(\Lambda_\qcd)$.
Specifically, we have computed the $W_i^\lat$ amplitudes in Feynman gauge.
In this case the weak boson propagator simplifies to the diagonal
form 
\begin{equation}
W_{\mu \nu} = \frac{ \delta_{\mu \nu} }{p^2 + \mw^2} 
\label{eq:wprop_feyn}
\end{equation}
but the amplitude, even at tree-level, requires also 
the contribution of the Goldstone boson arising after the 
Electro-Weak Symmetry Breaking.

\begin{figure}[htb]
\includegraphics[width=.3\textwidth]{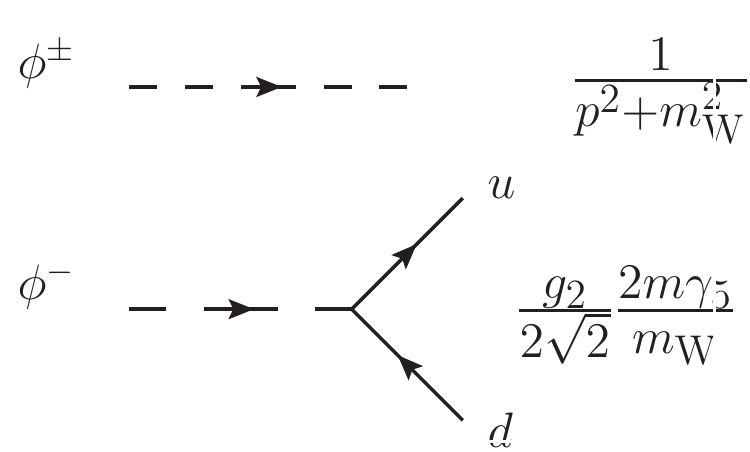}
\caption{Goldstone boson propagator and coupling to the 
quark current. In the latter we have already
assumed degenerate quark masses and simplified the contribution
proportional to the identity.}
\label{fig:gold}
\end{figure}

\begin{figure}[htb]
\includegraphics[width=.5\textwidth]{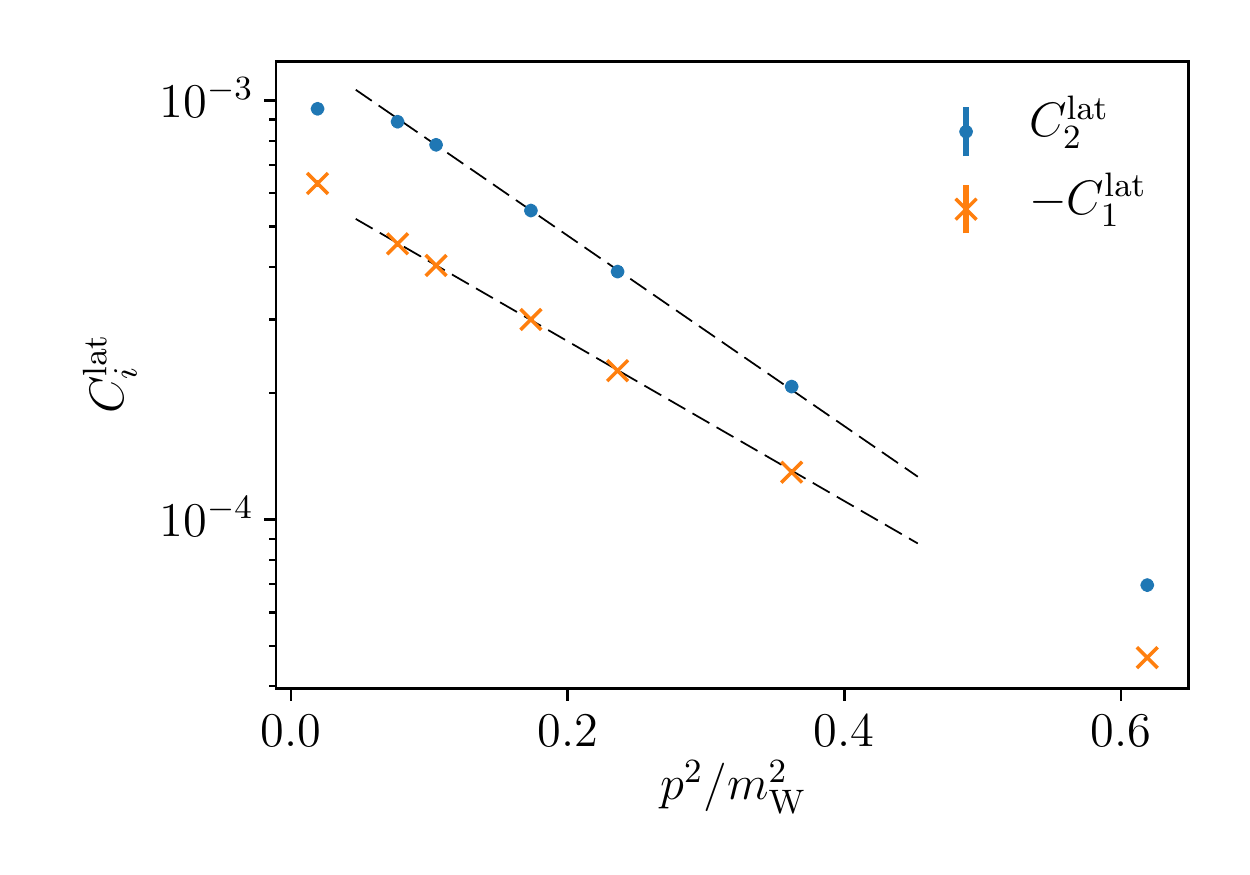}
\caption{Contribution of the Goldstone boson exchange 
to the Wilson coefficients with parity even 
$\gamma$ projectors. The two lines are linear fits of 
$\log C_i^\lat$ in $p^2/\mw^2$, which describe well the data.
The suppression given by the prefactor in the vertices, 
$(2m/\mw)^2 = 0.0064$, makes the two contributions negligible, even though
they slightly reduce the discrepancy from the $C_i^\lat$s 
obtained with parity odd projectors.}
\label{fig:c2-feyn}
\end{figure}

In our case, where all the external legs have the same mass, the 
vertex between the Goldstone and two quarks simplifies to the form
reported in \fig{fig:gold}.
In the limit of mass-less quarks it should vanish identically thus leaving
only \eq{eq:wprop_feyn} to contribute to the amplitude. 
This may be spoiled again by non-perturbative effects, which may be
very pronounced due to the $\gamma_5 \otimes \gamma_5$ structure
of this diagram.
Therefore studying the two contributions separately could shed 
some light on the size of these effects and we examine this in \fig{fig:c2-feyn},
which shows the Wilson coefficients $C_i^\lat$ computed only from 
the Goldstone boson diagram and with parity even projectors.
Note that the full Wilson coefficients are obtained simply by adding\footnote{
The multiplicative renormalization factor of this diagram is $Z_A$ 
(from the PCAC relation) and some care is required when considering the
renormalized amplitude $W_i^\ri$.}
the result of \fig{fig:c2-feyn}
to the $C_i^\lat$s computed from the $W$ exchange alone.

For parity odd projectors the contribution from the Goldstone boson
is negligible, numerically below $10^{-6}$ and compatible with
zero within 1 sigma.
For parity even projectors, in contrast, 
we can measure a signal from the Goldstone boson diagram. 
However the small prefactor $(2 m/\mw)^2$ makes this contribution
negligible also in this case, as plotted in \fig{fig:c2-feyn},
although it moves both $C_1^\lat$ and $C_2^\lat$ towards
the corresponding parity odd data.\\

Next we compare the Wilson coefficients measured in Unitary and 
Feynmann gauge with odd projectors exclusively, for which 
the Goldstone boson diagram can be neglected. 
The difference that we observe between the sets of data points 
comes from the gauge depedent part of the weak propagator in
\eq{eq:wprop} proportional to $p_\mu p_\nu$.
The dependence on the weak gauge fixing condition is expected
to vanish for on-shell quantities, which we approach by injecting smaller and smaller
momenta in the external quark legs of our amplitudes.
As seen already above, in this limit non-perturbative 
effects produce contaminations that in this case 
let gauge-dependent terms survive: results between Feynman and Unitary gauge
do not agree at the 2\% level, even for the preferred parity odd projectors, and 
we may consider this difference as one source of systematic uncertainties 
$\delta C_i^\mathrm{gauge}(\mw\,, p^2)= 
|C_i^{\lat,\mathrm{unit}}(\mw\,, p^2) - C_i^{\lat,\mathrm{feyn}}(\mw\,,p^2) | $.

However, a second systematic error that we include in our calculation is taken from
the difference between the even and odd projectors in Unitary gauge, 
$\delta C_i^\mathrm{proj} = |C_i^{\lat,\mathrm{VV+AA}} - C_i^{\lat,\mathrm{VA+AV}}|$ 
(for better readability we omit the $\mw$ and $p^2$ dependence).
The latter turns out to be much larger than $\delta C_i^\mathrm{gauge}$ 
and similarly accounts for non-perturbative
contaminations. Therefore we have decided to discard $\delta C_i^\mathrm{gauge}$.

Finally we also consider the quark mass dependence
as a source of systematic uncertainties, 
$\delta C_i^\mathrm{mass} = |C_i^{\lat,am=0.04} - C_i^{\lat,am=0.02}|$,
measured for exceptional kinematics only from parity odd projectors.\\

Now that we have established how to quantify remaining non-perturbative
effects, as our last step in the extraction of the Wilson coefficients we
study the dependence of the total systematic uncertainty on 
the lower end ($\pcut$) of the fit range of $C_i^\lat$ 
\begin{equation}
\delta C_i^\lat (\pcut) = \sqrt{(\delta C_i^\mathrm{proj})^2 + (\delta C_i^\mathrm{mass})^2 + 
(\delta C_i^\mathrm{poly})^2 } \,,
\label{eq:syst}
\end{equation}
where we also quote the extrapolation error $\delta C_i^\mathrm{poly}$ 
as a function of $\pcut$ by taking the difference of the 
intercepts of two different polynomial fits.

\begin{figure*}[ht]
\includegraphics[width=.49\textwidth]{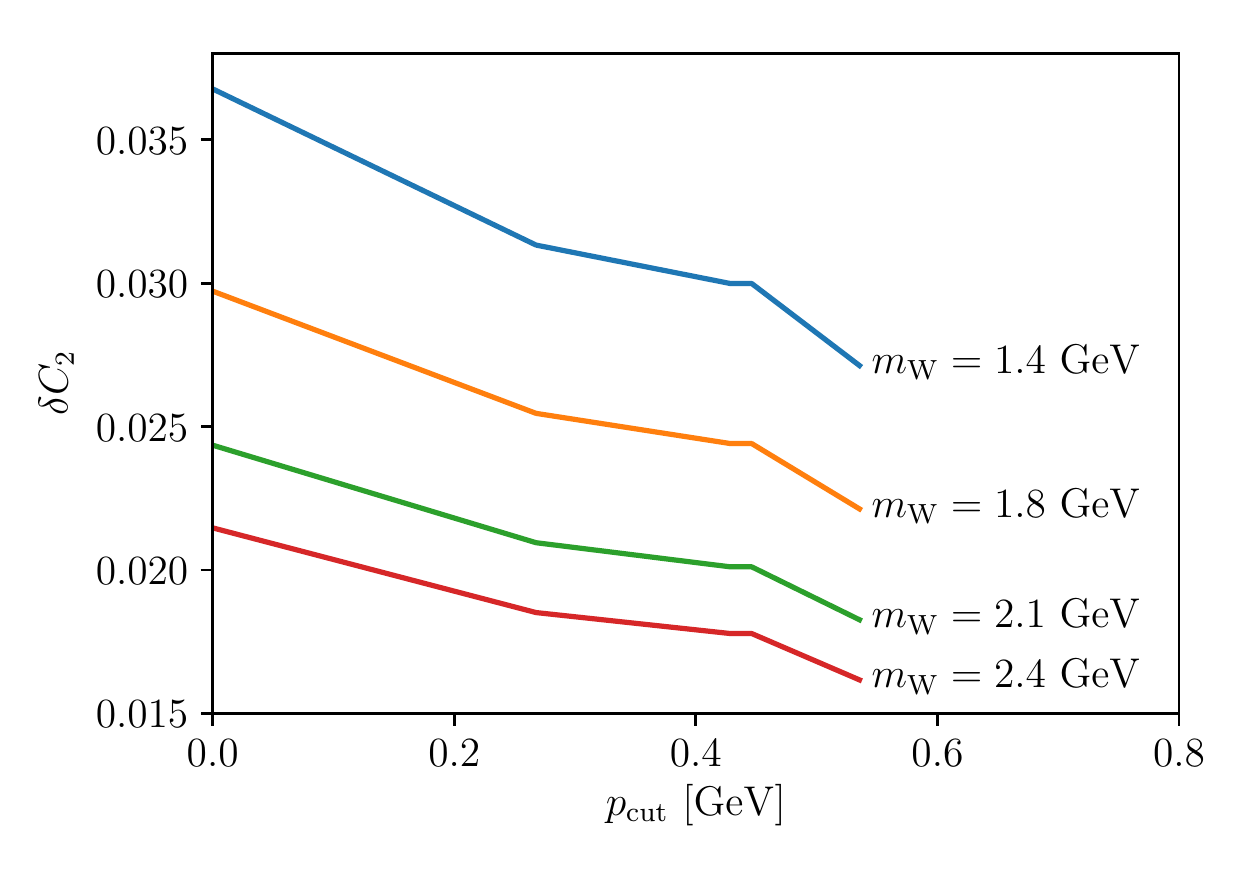}
\includegraphics[width=.49\textwidth]{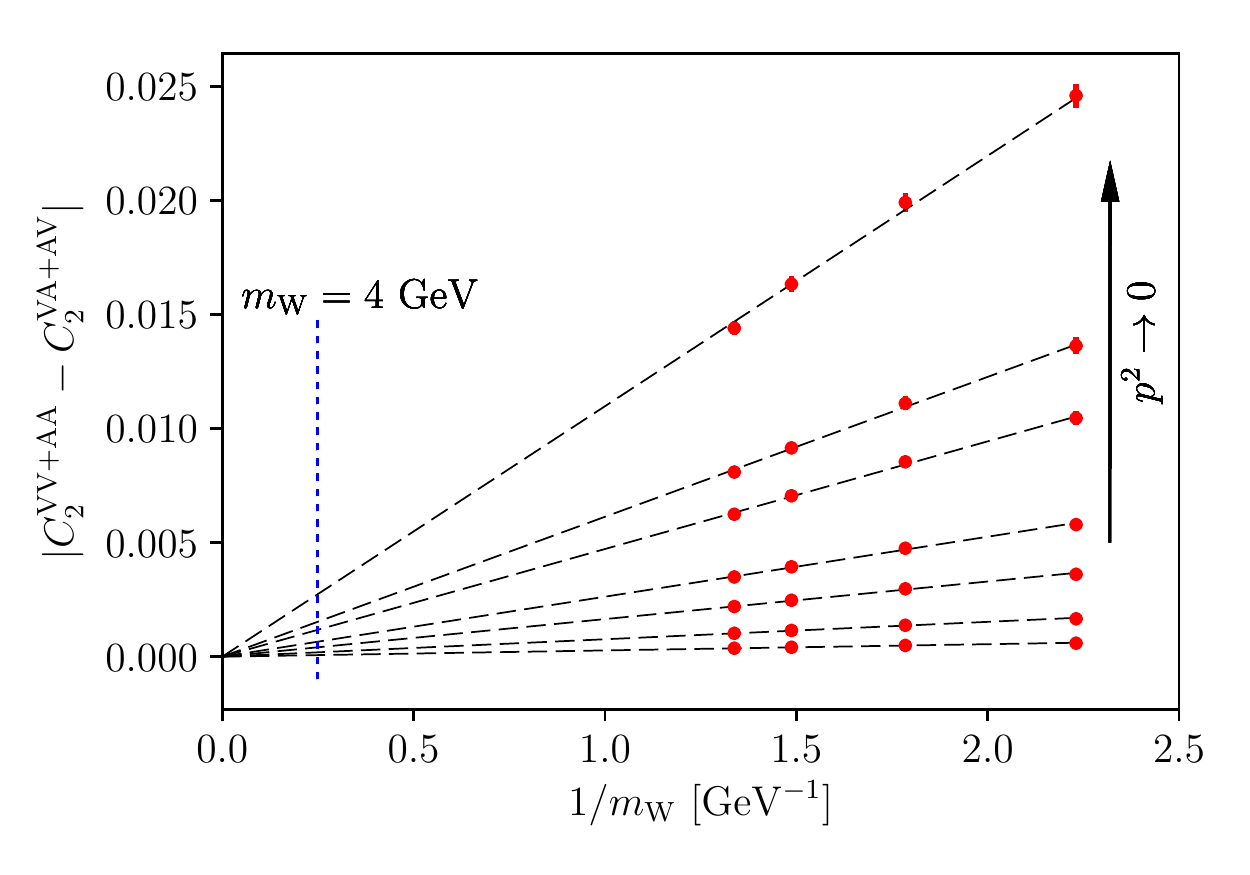}
\caption{{\it Left}: $\delta C_2$, as a function of $\pcut$, decreases 
as more points at small momenta are excluded from the fit, suppressing
non-perturbative effects. 
{\it Right}: measured difference of $C_2$ from odd and even projectors in the
$\gamma$ scheme as function of the $W$ boson mass. 
The lines are linear fits forced to pass through the origin
with excellent $\chi^2$. The non-perturbative contaminations which we
quantify with $\delta C_i^\mathrm{proj}$ given in \eq{eq:syst} seem to vanish only 
with one inverse power of $\mw$. In the plot we present the data 
for the exceptional kinematics, 
but practically no difference is observed in the non-exceptional case.}
\label{fig:syst}
\end{figure*}

In general, all the systematic uncertanties quoted in \eq{eq:syst} are estimated
from the combined fits defined in \eq{eq:fits} where we exclude the points
with $p^2 < \pcut^2$.
As we increase $\pcut$ in our fit ranges, $\Lambda_\qcd$-type contributions
are expected to vanish and this is reflected by our systematic uncertainty,
which decreases also for larger values of $\mw$. We demonstrate
both behaviors in the two panels of \fig{fig:syst}: the left one confirms
essentially the left-most inequality in \eq{eq:window} and the fact the
natural variable to control non-perturbative effects is precisely $\pcut$; 
the right plot also confirms the general expectation that uncertainties are
reduced with a larger separation between the $\qcd$ and weak scales.
An additional indication of the non-perturbative origin of these contributions
resides in the linear behavior in $1/\mw$ showed in \fig{fig:syst}: 
the presence of a condensate 
(a pure non-perturbative effect) 
could be captured by a higher dimensional operator
in the form of
\begin{equation}
\frac{ \bar q q } {p^2 \mw} O^{(6 \mbox{-} \mathrm{dim})} \,,
\end{equation}
whose functional form describes well the data.\\

To summarize, we have collected evidence, 
from the spread between different projectors and
gauge fixing conditions to the volume dependence,
that lead to infrared contaminations vanishing with $1/\mw$.
In the right plot of \fig{fig:syst} we also demonstrate how a future
calculation with $\mw \approx 4~\GeV$ would significantly
improve the precision well below the percent level.

Finally, for the central values of the Wilson coefficients
we use a $\pcut$ of $0.3~\GeV$ for the 24I ensemble and $0.24~\GeV$ for the 32I
ensemble and we perform cubic and quadratic fits respectively
according to \eq{eq:fits}. We use amplitudes computed in unitary
gauge projected on the odd sector with $\qsl$ type projectors.
No significant effect, beyond the ones already considered, is obtained
from the difference with $\gamma$ projectors.

\section{Non-perturbative determination of the Wilson coefficients}

The remaining systematic uncertanties that we need to address are
related to discretization errors for which we need the renormalization
factors. In the following we present non-perturbative results 
for the following values of the $W$ boson mass: 1.4, 1.8, 2.1 and 2.4 $\GeV$.

\subsection{Renormalization factors}

As before twisted boundary conditions turn out to be very useful, as 
they give us the possibility to tune the momentum to the desidered value.
In our computation of the renormalization factors we have adopted the conventional
RI/SMOM scheme described in \Ref{Lehner:2011fz}.
Here a momentum $2p$ leaves the operator, as opposed to \eq{eq:nonex},
and only two inversions per configuration are
required with momentum $p_1=(x,-x,0,0)$ and $p_2=(0,x,-x,0)$ (with $p_3=p_1$ and $p_4=p_2$).
Since the renormalization conditions in \eq{eq:rimom} are imposed in the chiral limit
we repeated the measurements for two values of the quark mass ($am=0.04$ and 0.02) 
and we extrapolate linearly to zero quark masses.
We computed the quark propagators such that $ |ap| \approx a \mw$, 
to obtain the Wilson coefficients renormalized at a scale coinciding with the mass of 
the weak boson utilized in their definition: in this way we are reproducing the 
so-called initial conditions, where large logarithms are avoided as explained in
section II.

Finally, we deal with the renormalization of the 
local currents on the lattice.
Therefore from the same set of propagators we also compute the following quantities
\begin{equation}
\begin{split}
\Gamma^V_\mu = & \sum_x \langle \gamma_5 \tilde G(x,-p_2)^\dagger \gamma_5 \ \gamma_\mu \ \tilde G(x,p_1) \rangle \,, \\
\Gamma^A_\mu = & \sum_x \langle \gamma_5 \tilde G(x,-p_2)^\dagger \gamma_5 \ \gamma_\mu \gamma_5 \ \tilde G(x,p_1) \rangle \,.
\end{split}
\label{eq:GammaV_A}
\end{equation}
After the usual amputation with the inverse propagators, we impose the renormalization conditions
\begin{align}
\lim_{m_q \to 0} \frac{Z_V}{Z_q^\ri} \Tr[ \Lambda_\mu^V \gamma^\mu ] = \Tr [ \Lambda_\mu^{V,\tree} \gamma^\mu] \,, 
\label{eq:ZV} \\
\lim_{m_q \to 0} \frac{Z_A}{Z_q^\ri} \Tr [\Lambda_\mu^A \gamma^\mu \gamma_5 ] = \Tr [ \Lambda_\mu^{A,\tree} \gamma^\mu \gamma_5 ] \,. 
\label{eq:ZA}
\end{align}
In principle the appropriate renormalization condition should involve
 the $\Gamma^{V-A}_\mu$, obtained by replacing $\gamma_\mu$ with $\gamma_\mu^L$ in the first
line of \eq{eq:GammaV_A}. However we have explicitly checked from our measurements that 
$\Tr[ \Lambda_\mu^A \gamma^\mu] \ll 10^{-3}$ (the same holds if $A \to V$ and 
$\gamma^\mu \to \gamma^\mu \gamma_5$) which means 
that for the choice of projectors in eqs.~(\ref{eq:ZV},\ref{eq:ZA})
only $Z_V$ alone (or $Z_A$) renormalizes the weak coupling $g_2$, 
thus leading to \eq{eq:matching}.
Replacing $\gamma_\mu \to (\qsl q_\mu)/q^2$ and $\gamma_\mu \gamma_5 \to (\qsl \gamma_5 q_\mu)/q^2$
in eqs.~(\ref{eq:ZV},\ref{eq:ZA})
defines the $\qsl$ scheme as before. 
For simplicity, we do not 
combine
$Z_q$ and the four-quark $Z$ matrix from two schemes.\\

Let us briefly examine the properties of the renormalization
factor of the Wilson coefficients in \eq{eq:wcoeff}
\begin{equation}
\tilde Z^{\ms,\ri} \equiv Z_V^2 [Z^\ri]^{-1} [Z^{\ri \to \ms}]^{-1} \,,
\end{equation}
which we explicitly expand in terms of lattice observables for the
reader's convenience (in the $\gamma$ scheme)
\begin{equation}
Z_V^2 [Z^\ri]^{-1}_{ik} = \left( \frac{\Tr [ \Lambda_\mu^{V,\tree} \gamma^\mu] }{\Tr[ \Lambda_\mu^V \gamma^\mu ]} \right)^2
\cdot M^\lat_{ij} [M^\tree]^{-1}_{jk} \,.
\end{equation}

Firstly the quark mass dependence is negligible, as the four
elements of the matrix $\tilde Z$ slightly differ when the quark
mass is changed from 0.04 to 0.02. Nevertheless we take this
into account by linearly extrapolating to the chiral limit.
Secondly we examine the difference between the renormalization 
matrix $\tilde Z$ computed from two intermediate schemes, 
defined by the two classes of projectors, $\gamma$ and $\qsl$.
As we change the renormalization scale from 1.4 to 2.4 $\GeV$
the conversion factor $Z^{\ri \to \ms}$, which is only perturbative, 
becomes more precise. 
Nonetheless 
the ratio $\tilde Z^{\ms, \gamma} [\tilde Z^{\ms,\qsl}]^{-1}$
significantly deviates from the identity matrix,
the signature of missing terms of size $\alpha_s^2$.
The largest departures from 1 are observed in the diagonal terms,
specifically from 23 to 18\% for the $(1,1)$ element 
(that contributes mostly to $C_1^\ms$), 
and from 5 to 4\% for the $(2,2)$ element that defines
$C_2^\ms$. These higher order effects are relatively large and
can be reduced only by pushing the calculation to higher 
renormalization scales, where $\alpha_s$ becomes smaller.
Thirdly, at these relatively high momenta, 
both parity even and odd projectors give results
well compatible with each other.

\subsection{Discretization Errors}

\begin{figure}[ht]
\includegraphics[width=.49\textwidth]{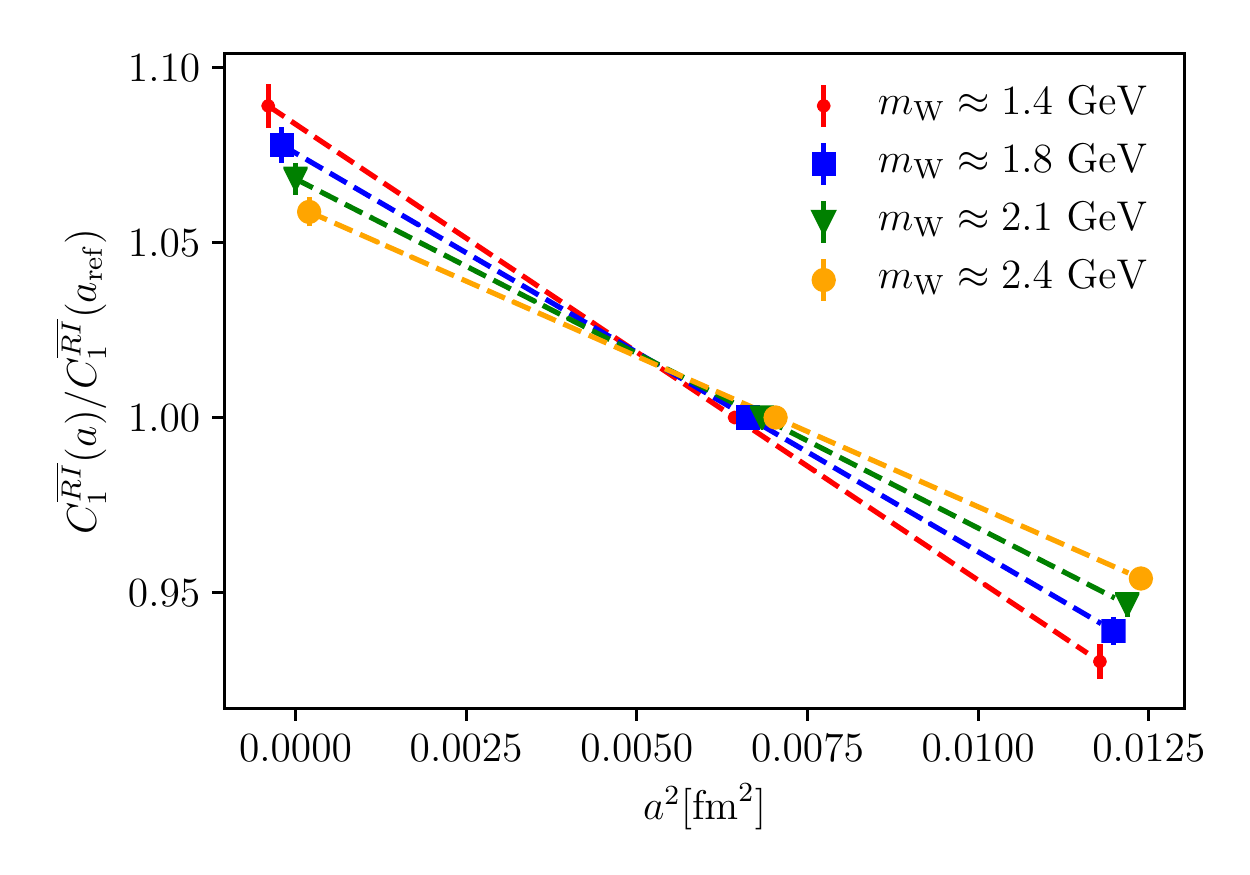}
\caption{Relative continuum extrapolations of $C_1^\ri$, 
renormalized at the value of the $W$ boson mass used to compute them. 
We normalize the y-axis with the measurement of $C_1^\ri$ from 32I. 
The slope of the extrapolations slightly changes with $\mw$, but remains
a subdominant effect. The points have been horizontally 
displaced for better visibility.}
\label{fig:cutoff}
\end{figure}

The lattice cutoff is the main limitation of the current results preventing 
the weak boson mass from being well separated from lower $\qcd$ scales. 
As we increase it, larger discretization errors are expected and values of 
$\mw$ of the order of the cutoff may sound already dangerous:
in \fig{fig:cutoff} we demonstrate that the parameter space 
explored in this work is still in a region where discretization errors
are reasonably under control.

By plotting the continuum extrapolations of $C_1^\ri$
for the four values of $\mw$ considered and 
normalized at the finer ensemble 32I, 
we show how the size of the $a^2$ coefficient slightly changes with 
$\mw$, making the $\mw$-dependence practically negligible 
compared to the overall size of cutoff effects. 
The scaling violations that we observe for $C_1^\ri$ 
range from 10 to 17\%; instead they are only at 
1\% level for $C_2^\ri$, where
the differences among the several values of $\mw$
are also irrelevant.
The different magnitudes
can be easily understood
from the fact that in the free theory $C_1=0$ and $C_2=1$, 
meaning that discretization errors 
start at order $\alpha_s$,
which on our lattices is approximately 0.3.
Nevertheless we perform extrapolations in $a^2$ 
rather than $\alpha_s(a) a^2$, since the numerical 
change is irrelevant.

\subsection{Final results and discussions}

\begin{figure*}[ht]
\includegraphics[width=.49\textwidth]{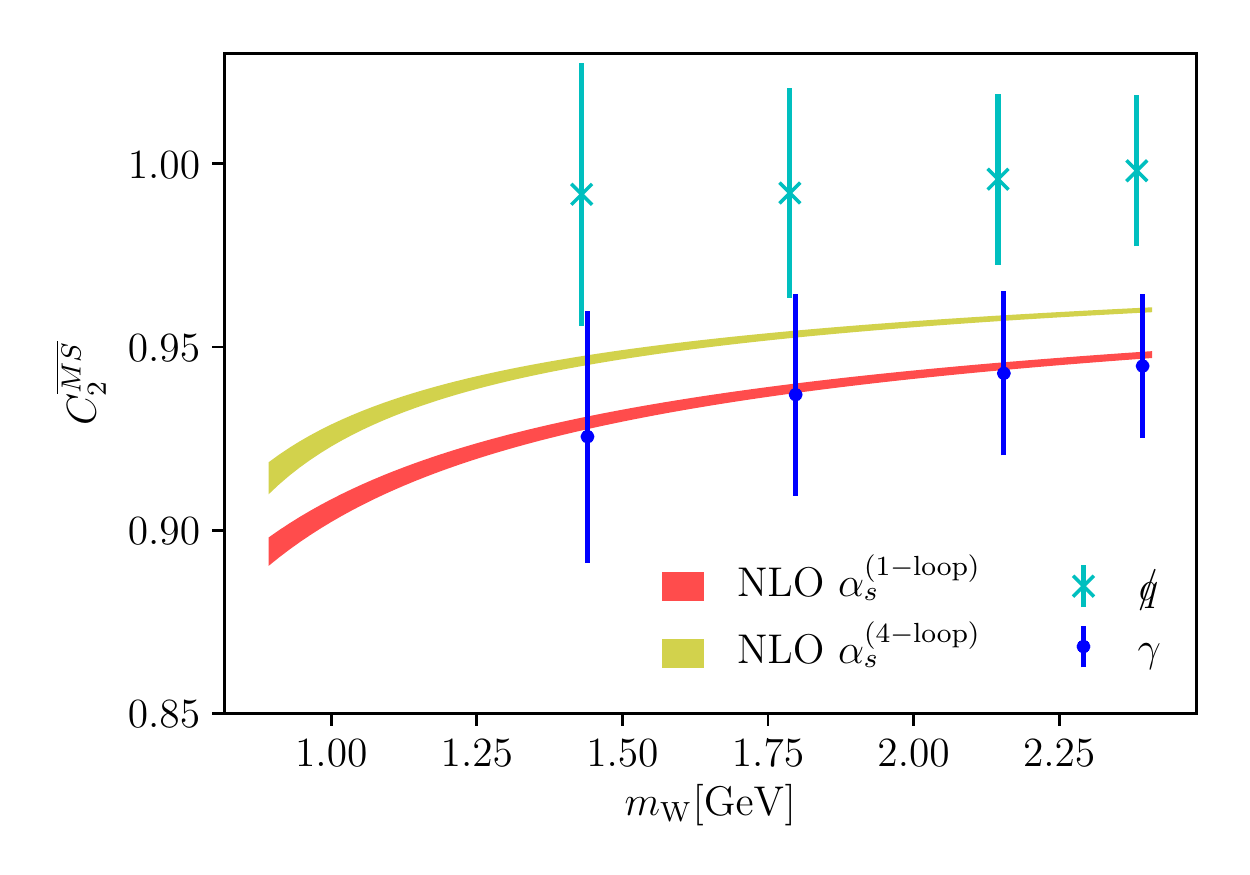}
\includegraphics[width=.49\textwidth]{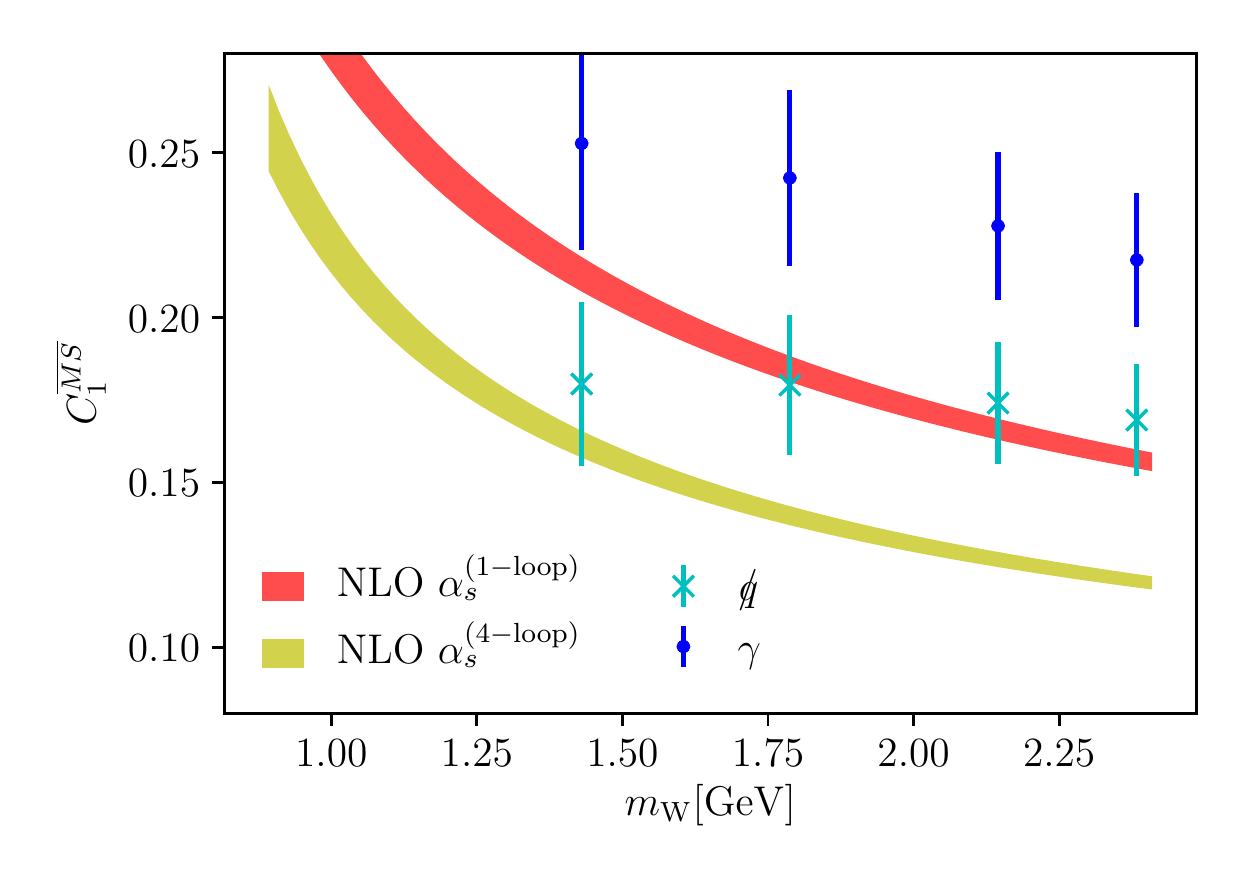}
\caption{Results in the continuum limit for 3-flavor QCD 
of the Wilson coefficients in the 
$\ms$ scheme as a function of the $W$ boson mass.
The shaded regions represent the perturbative one-loop result with error
propagated from the $\Lambda$-parameter. The difference between the two intermediate
$\ri$ schemes accounts for higher order effects in the $\ri \to \ms$ conversion
factors. The plotted error bars include both statistical and systematic uncertainties.
The latter dominate and gradually decrease with $\mw$.}
\label{fig:c12ms}
\end{figure*}

The last aspect of our work consists of changing renormalization scheme from $\ri$ to the 
more common $\ms$. This is the only step where perturbation theory enters in our calculation
since the conversion matrices are known to 1-loop and reported in \app{app:rims}. 
In \fig{fig:c12ms} we plot our results together with the known 1-loop analytic formulae that we 
take from \Ref{Buchalla:1995vs}. For the latter we use $\alpha_s$ from the 1 and 4-loop $\beta$
function with 3 flavors and $\Lambda^\ms_{N_\mathrm f =3} = 341(12) \MeV$ taken from \Ref{Bruno:2017gxd}.
The error from the $\Lambda$ parameter is propagated to the analytic 
Wilson coefficients and it is represented by the two shaded regions.

The non-perturbative results include the various systematic errors described
in the previous sections. The systematic uncertainty associated with the
perturbative error of the $\ri \to \ms$ conversion, which is an $O(\alpha_s^2)$ effect,
can be read off from the difference between
the two intermediate $\ri$ schemes studied in the paper, the $\gamma$ and $\qsl$ scheme.
Such a difference turns out to be relatively large, following the previous
discussion on the renormalization matrix $\tilde Z$.
We remark that the $\ri \to \ms$ conversion matrix for the 
$\qsl$ scheme given in \app{app:rims} contains a large one-loop coefficient,
whereas the $\gamma$ scheme shows a much better convergence with a very small correction
from $\ri$ to $\ms$.

In \fig{fig:c12ms} we compare our results for the $C_i^\ms$s against the perturbative
ones as a function of the $W$ mass. Recall that we are focusing on the initial conditions of
the Wilson coefficients, which means that their renormalization scale coincides with $\mw$.
The discrepancy between our values and the 1-loop results decreases for larger values of
the weak mass as expected from the running of the strong coupling constant, 
since the perturbative expansion of the Wilson coefficients reads
\begin{equation}
C_i(\mw) = \sum_{n=0}^N \alpha_s^n (\mw) k^{(i)}_n + O(\alpha_s^{N+1})
\label{eq:wcoeff_pert}
\end{equation}
with tree-level values $k_0^{(1)}=0$ and $k_0^{(2)}=1$.
The systematic uncertainties plotted in \fig{fig:c12ms} 
are all correlated with each other, allowing us to fit the data with 
\eq{eq:wcoeff_pert} and still be able to estimate the one-loop coefficients:
from the $\gamma$ intermediate scheme alone we obtain 
$k_1^{(1)}=0.64(21)$ and $k_1^{(2)}=-0.158(78)$, that agree within 1 sigma
with the analytic results taken from \Ref{Buchalla:1995vs}, 0.44 and -0.15 
respectively\footnote{The spread introduced by different definitions 
of $\alpha_s$ is included in the quoted errors of the
1-loop coefficients}.\\

Given the limitations of the $\ri \to \ms$ conversion factor to one-loop, even with higher
precision, fitting our $\ms$ results beyond $O(\alpha_s)$ would be incorrect.
In fact, we turn now to our non-perturbative determination by sticking to the 
$\ri$ scheme, where we can convert the analytic results of the Wilson coefficients
without loss of generality.
In \tab{tab:c12fits} we report various results from different polynomial fits to the 
$\gamma$ scheme, following \eq{eq:wcoeff_pert}, which we also 
augumented with a term $k_\Lambda /\mw$.

\begin{table}[ht]
\begin{tabular}{cccc}
\toprule
$C_i^\ri$ & $k_1^{(i)}$ & $k_2^{(i)}$ & $k_\Lambda^{(i)}$ \\
%\midrule
\colrule
1 & 0.643(60)  & --- & --- \\
1 & 1.270(45)  & -1.28(0.18) & --- \\
1 & 0.123(13)  & --- & 0.337(91)    \\
\colrule
2 & -0.158(61) & --- & --- \\
2 & -0.223(24)  & 0.01(0.28)  & --- \\
2 & -0.006(22) & --- & -0.14(11)  \\
%\bottomrule
\botrule
\end{tabular}
\caption{In the first lines we report results for the Wilson coefficient $C_1$, 
whereas the last ones refer to $C_2$. All fits have very small $\chi^2$ per $\mathrm{d.o.f.}$
due to the large systematic errors.
The errors quoted in the table are both statitstical and systematic.
In our fits we always use all the four values of $\mw$ explored in this work.
}
\label{tab:c12fits}
\end{table}

The coefficients reported in \tab{tab:c12fits} could be used, in principle, 
to provide an estimate of the Wilson coefficients for physical values
of $\mw$ by using \eq{eq:wcoeff_pert} with $\alpha_s$ computed 
at $80~\GeV$. 
With the current large systematic uncertainty we are only sensitive to the 1-loop 
coefficient, but the results in \tab{tab:c12fits} show the potential of the method:
the possibility to extract higher loop coefficients can
 be used to bound the error of $C_i(80~\GeV)$ by considering, for example, the
 difference between the known 1-loop result and our 2-loop prediction.
This approach has the potential to reduce the current systematic uncertainties
where these Wilson coefficients are used and are relevant, 
such as the real part of the amplitudes 
of $K \to \pi \pi$ for isospin 0 and 2 channels.
One caveat that we need to remember is that any prediction of coefficients
beyond one-loop depends on the number of flavors used in the simulations.
For this reason we intend to continue our study by reusing the methodology
developed in this paper 
on the finer ensembles generated by the RBC/UKQCD 
Collaboration \cite{Frison:2014esa} with 3 and 4 active flavors in the sea.
Such a calculation will provide multiple benefits, from the possibility to
push $\mw$ up to 4 $\GeV$ and substantially reduce the systematic uncertainties
 (see \fig{fig:syst}), to controlling the flavor dependence of the coefficients of 
 higher loops and reducing the systematic error from intermediate $\ri$ 
 scheme, thus providing a solid prediction in the $\ms$ scheme.
Finally, future extensions of this work will include the top quark contribution.

\section{Conclusions}

In this paper we have presented a method to compute the initial conditions of the
Wilson coefficients to all orders in the strong coupling constant and leading order
in $g_2$. We have described the limitations of our exploratory study, mostly 
related to the presence of large infrared and non-perturbative effects.
By looking at different observables we quantified those effects and took them into
accout in our systematic uncertainties, that dominate the final errors.

Nevertheless, we have demonstrated how precise statistical results can be achieved
with the combined use of momentum sources and twisted boundary conditions. Therefore
we expect to obtain excellent results with the next iteration of this calculation repeated
on finer lattices, where the systematic errors will significantly decrease.

Despite the limitations imposed by the lattice cutoff, we observed reasonable scaling
violations for the values of $\mw$ that we explored. Moreover we discussed 
a strategy to extend the relevance of our study to place a bound on the error
of the Wilson coefficients for the physical scenario with $\mw \approx 80 ~\GeV$.

\section{acknowledgements}

We would like to thank our RBC and UKQCD collaborators for helpful discussions and support.
M.B. is particularly indebeted to N.~Christ for a critical reading of the manuscript
and to T.~Izubuchi and P.~Boyle for many stimulating discussions.
Computations for this work were carried out at the Fermilab cluster pi0 
as part of the USQCD Collaboration,
which is funded by the Office of Science of the U.S.
Department of Energy.
M.B., C.L. and A.S. were supported by the United States 
Department of Energy under Grant No. DE-SC0012704.
In addition C.L. is supported in part 
through a DOE Office of Science Early Career Award.

\appendix
\section{Stochastic sampling of the weak propagator} 
\label{app:point-source}

In this Appendix we present further details on the alternative 
stochastic sampling method for the $W$ boson propagator described in the main text.
The key ingredient is the possibility to approximate the propagator in momentum
space 
$\sum_x e^{ip(y_0-x)} D^{-1}(x,y_0)$
with point-source propagators transformed to any
continuous momentum, up to finite-size errors
\begin{equation}
\begin{split}
& S^\mathrm{app} (p) = \sum_x D^{-1}(x,y_0) \\ 
& \times \left \lbrace 
\begin{array}{ll}
e^{ip_\mu (y_0-x)_\mu} & |y_0-x|_\mu<L_\mu/2 \\
e^{ip_\mu (y_0-x + L)_\mu} & (y_0-x)_\mu \leq -L_\mu/2 \\
e^{ip_\mu (y_0-x - L)_\mu} & (y_0-x)_\mu \geq L_\mu/2  \\
\end{array} \right.
\end{split}
\end{equation}
It is easy to check that if the momentum is an allowed fourier mode
the equation above amounts to a simple translation of the source to the origin. 
However if the momentum is not quantized,
$S_\mathrm{app}$ approximates well the propagator in infinte volume:
the mass gap of QCD ensures that finite volume errors are exponentially
small.\\

The second feature of the stochastic sampling relies on the approximation
of the sum over the $W$ propagator. Let us fix one end to the origin
and call $r$ the second end.
Due to the fast fall-off of the weak boson
propagator, only small separations 
contribute to the signal. Hence, we start by dividing all distances $r$
in classes defined by their absolute value $|r|$ with multiplicities 
$d_{|r|}$. Then we randomly
choose one representative per class and we cover all distances 
up to a certain cutoff $R_\mathrm{inner}$. Up to lattice symmetries
this sum is exact.

Finally we sample the remaining classes
starting from $R_\mathrm{inner}$ with probability
$p(r) \propto |r|^{-3}$ and we evaluate the appropriate reweighting 
probability $w(r)$ to obtain the flat sum again.
With $a \mw=1.0$ on the 24I ensemble, sampling 30 classes
exactly (below $R_\mathrm{inner}$) and 10 classes stochastically 
(above $R_\mathrm{inner}$) produces controlled approximations
 with stochastic errors around 1\%.
This sampling strategy defines what we call a ``cloud'' of points around the
origin.

As explained already in the text, the problem resides in the second sum 
over different clouds, each centered around a randomly chosen point. 
Although the noise of the final amplitudes and Wilson coefficients
scales with the number of clouds, compared to the momentum sources, 
which amount to summing all of them,
it is still from 5 to 10 times larger, when going from high ($1~\GeV$) to 
small momenta.
The access to all momenta has the additional advantage of averaging
over different orientations, such as $(p,0,0,0)$, $(0,p,0,0)$, $(-p,0,0,0)$, 
etc\dots, an effect already included in the numerical factors quote above.\\ 

To reduce the cost of this method we have also employed an All-Mode-Averaging~\cite{Blum:2012uh}
strategy by computing the point-source propagators of the clouds 
sloppily and adding the corresponding correction term, 
which we tuned to be well below the statistical error
throughout the entire range of momenta. In our final measurements
we have used up to 16 total clouds, each containing 40 points as described above.
For the same cost we have obtained the data points showed in \fig{fig:c12lat}
and in the right panel of \fig{fig:finite_volume}.

\section{$\ri \to \ms$ conversion matrices}
\label{app:rims}
The $Z^{\ri \to \ms}$ for the $\gamma$ scheme for $Q_1$ and $Q_2$ in
the RI/MOM scheme can be found in \Ref{Buras:2000if}. 
We have chosen instead the RI/SMOM scheme, whose conversion factors to the $\ms$ scheme
are known in the 3-flavor EFT for the 10 operator 
basis of $\Delta S=1$ transitions
both for $\gamma$ and $\qsl$ projectors~\cite{Lehner:2011fz}.
Below we show how to derive the conversion matrix for $Q_1$ and $Q_2$ 
in the four-flavor theory from the three-flavor results of \Ref{Lehner:2011fz}.

We start from the three-flavor EFT and we restrict ourselves to the first three
operators of basis I of \Ref{Lehner:2011fz} (the superscript $^{(3)}$ 
distinguishes the following operators from $Q_1$ and $Q_2$ introduced in 
\eq{eq:Q1_Q2}):
\begin{equation}
\begin{split}
Q_1^{(3)} = & (\bar s_i \gamma_\mu^L u_j) (\bar u_j \gamma_\mu^L d_i) \,, \\
Q_2^{(3)} = & (\bar s_i \gamma_\mu^L u_i) (\bar u_j \gamma_\mu^L d_j) \,, \\
Q_3^{(3)} = & (\bar s_i \gamma_\mu^L d_i) \sum_q (\bar q_j \gamma_\mu^L q_j) \,. 
\end{split} 
\label{eq:Q123}
\end{equation}
Then we substitute $u$ with $c$ whenever it appears in \eq{eq:Q123}
and we compute the Green's functions, similarly to \eq{eq:greenfunc},
\begin{equation}
\begin{split}
[\Gamma(Q_i^{(3)})]^{\alpha \beta \gamma \delta}_{a b c d}   = 
\langle \, s^\alpha_a u^\gamma_c \, Q_i^{(3)} \, \bar c^\beta_b \bar d^\delta_d \, \rangle \,.
\end{split}
\end{equation}
With the substitution $u \to c$ advocated above, 
we are explicitly eliminating all the disconnected diagrams 
of the three-flavor theory. At this point
it is easy to demonstrate that
\begin{equation}
\begin{split}
\Gamma(Q_1^{(3)}) = \, & \Gamma(Q_1) \,, \\ 
\Gamma(Q_2^{(3)}) = \, & \Gamma(Q_2) \,, \\ 
\Gamma(Q_3^{(3)}) = \, & \Gamma(Q_1) \,,
\end{split}
\label{eq:Q123_Q12}
\end{equation}
where the operators on the r.h.s. correspond to \eq{eq:Q1_Q2}.
In the three-flavor theory the 10 operator basis is redundant, 
leading to a smaller basis, with linear independent operators,
often called the chiral basis (we consider again solely the linear
combinations involving the three operators in \eq{eq:Q123})
\begin{equation}
\begin{split}
Q_1^\prime = & 3 Q_1^{(3)} + 2 Q_2^{(3)} - Q_3^{(3)} \,, \\
Q_2^\prime = & \tfrac{1}{5} (2 Q_1^{(3)} - 2Q_2^{(3)} + Q_3^{(3)}) \,, \\
Q_3^\prime = & \tfrac{1}{5} (-3 Q_1^{(3)} +3 Q_2^{(3)} + Q_3^{(3)}) \,.
\end{split}
\label{eq:chiral_basis}
\end{equation}
If we now combine \eq{eq:chiral_basis} with \eq{eq:Q123_Q12} we can relate the chiral
basis to our two operators as follows
\begin{equation}
\Gamma(Q_i^\prime) = R_{ij} \Gamma(Q_j) \,,  \quad 
R_{ij} = \begin{pmatrix}
2 & 2 \\
3/5 & -2/5 \\
-2/5 & 3/5
\end{pmatrix} \,.
\end{equation}
In \Ref{Lehner:2011fz} the conversion matrices $Z^{\ri \to \ms}$ 
are given in the chiral basis, which is the reason behind 
the introduction of the matrix $R$.
We stress that $Q_1^\prime$ transforms under the (27,1) representation
of the chiral group, whereas $Q_2^\prime$ and $Q_3^\prime$ under
the (8,1) representation. This prevents any mixing between the two sectors
once the penguin operators are discarded, as in our case.
Finally let us introduce three ad hoc matrices $T^{(i)}$ 
\begin{equation}
\begin{split}
T^{(1)} = & \begin{pmatrix}
1/5 & 1 & 0 \\
3/10 & -1 & 0
\end{pmatrix} \,, \\
T^{(2)} = & \begin{pmatrix}
0 & 3 & 2 \\
0 & 2 & 3 \\
\end{pmatrix} \,, \\
T^{(3)} = & \begin{pmatrix}
 3/10 & 0 & -1 \\
  1/5 & 0 & 1 \\
\end{pmatrix} \,,
\end{split}
\end{equation}
such that
\begin{equation}
  T^{(i)} R = \begin{pmatrix}
  1 & 0 \\
  0 & 1 \\
\end{pmatrix} \,, \quad
\forall i \,.
\end{equation}

To relate the renormalization factors of \Ref{Lehner:2011fz} to 
our specific case we need to recall the renormalization conditions 
in \eq{eq:rimom} and the definition of the matrix $M$. Similarly to \eq{eq:M} 
we can define a matrix $M^\prime$ for the chiral basis
\begin{equation}
M^\prime_{ij} = \Tr( P^\prime_j \Gamma(Q^\prime_i) ) = 
(R M U^T)_{ij} \,,
\label{eq:Mprime}
\end{equation}
with the matrix $U$ being the rotation of the projectors $P^\prime_i = U_{ij} P_j$.
Similar to the matrices $T^{(i)}$, we introduce three ad hoc matrices $S^{(i)}$ such that
their product with $U$ returns the $2\times 2$ identity.

Starting from the usual renormalization conditions for $M^\prime$ 
in the $\ri$ sense, with the help of \eq{eq:Mprime} and a few algebraic steps
we finally obtain
\begin{equation}
Z^{\ri \to \ms}_{2 \times 2} = T^{(i)} \cdot Z_{3 \times 3}^{\ri \to \ms, \prime} \cdot R  
\label{eq:Zri2ms}
\end{equation}
The universality of $\gamma$ scheme is such that for any choice of $U$ (and consequently of 
$S^{(i)}$) \eq{eq:Zri2ms} is always valid for $i=1,2$ and 3 and we verified that.
Instead for $\qsl$ projectors the situation is different, since a naive choice
of projectors can lead to mixing between the (27,1) and (8,1) operators even
with the explicit suppression of the penguin diagrams.
An accurate choice of projectors that forbids this is provided in the Appendix B of 
\Ref{Lehner:2011fz}. For simplicity we consider \eq{eq:Zri2ms} for $T^{(2)}$ 
which corresponds to selecting $Q_2^\prime$ and $Q_3^\prime$ alone.

The final results for the conversion matrices reported below have been obtained
from Table III of \Ref{Lehner:2011fz} for the $\gamma$ scheme and
from Table XIII for the $\qsl$ scheme where in both cases we have 
set the penguin contributions to zero
\begin{equation}
Z^{\ri \to \ms} = \mathbf{1}_{2 \times 2} + \frac{\alpha_s}{4 \pi} \Delta r \,,
\end{equation}

\begin{equation}
\Delta r_{\gamma, \gamma} = \begin{pmatrix}
\frac{8}{N_c} - \frac{12 \log 2}{N_c} & -8 + 12 \log 2 \\
-8 + 12 \log 2 & \frac{8}{N_c} - \frac{12 \log 2}{N_c} \\
\end{pmatrix} \,,
\end{equation}

\begin{equation}
\Delta r_{\qsl, \qsl} = \begin{pmatrix}
\frac{9}{N_c} - \frac{12 \log 2}{N_c} & -9 + 12 \log 2 \\
-7 + 12 \log 2 & \frac{9}{N_c} - 2 N_c -\frac{12 \log 2}{N_c} \\
\end{pmatrix} \,.
\end{equation}

\section{Subtracted projectors}
\label{app:subproj}

As explained in the main text, the bare lattice Wilson coefficients
can be obtained from any reasonable choice of projectors in the 
definition of $M$ and $W$.
Below we present an alternative definition of the projectors defined
in \eq{eq:gammaproj} which holds both for $\gamma$ and $\qsl$ types.
Since the aim of this digression is to reduce the discrepancy of the 
results obtained in the parity even and odd sectors, we focus
on the former, the most problematic one.
Let us introduce the ``subtracted projectors'' 
\begin{equation}
\tilde P_i = P_i^\mathrm{VV+AA} + b_{ij}^\alpha P_j^\alpha \,,
\end{equation}
with $i,j=1,2$ referring as usual to the color diagonal and mixed case
and the index $\alpha$ labelling the wrong chiralities that could potentially
mix in the parity even sector, namely $\mathrm{SS} \pm \mathrm{PP}$,
$\mathrm{VV}-\mathrm{AA}$ and $\mathrm{TT}$. 
The coefficients $b_{ij}^\alpha$ are fixed by solving the following
system 
\begin{equation}
\begin{split}
&\Tr ( \tilde P_i \, Q_j^\alpha ) = 0 \,, \\
\alpha=&\, \mathrm{SS}  \pm \mathrm{PP}, \mathrm{VV}-\mathrm{AA}, \mathrm{TT} \,,
\end{split}
\end{equation}
which minimizes the overlap of the projectors on the operators with the 
corresponding wrong chiralities (the index $i$ in the operator refers 
again to the color structure).
From a numerical experiment we found that the Dirac structures
$\mathrm{SS} - \mathrm{PP}$ and $\mathrm{VV}-\mathrm{AA}$ contribute
only to the noise without producing any effect.
Instead the remaining two chiralities have a beneficial effect
on the Wilson coefficients: in particular the $\mathrm{SS} + \mathrm{PP}$
alone reduces the discrepancy for both $C_1^\lat$ and $C_2^\lat$ by 20-30 \%; 
the inclusion of the tensor structure further increases the agreement 
with the odd projectors to less than 1 sigma for $C_2^\lat$, but 
has the opposite effect on $C_1^\lat$. In conclusion, the best result 
for the subtracted even projectors has been obtained with 
$\alpha = \mathrm{SS} + \mathrm{PP}$ alone. Since the benefit was marginal, 
in our estimate of $\delta C_i^\mathrm{proj}$
we decided
to use the un-subtracted projectors,
giving a more conservative and safer error.

\bibliographystyle{apsrev4-1}
\bibliography{wcoeff}

\end{document}